\documentclass[%
 reprint,
 amsmath,amssymb,
 aps,
]{revtex4-2}

\usepackage{graphicx}
\usepackage{dcolumn}
\usepackage{bm}
\usepackage{amsmath}
\usepackage{mathtools}
\usepackage{float}
\makeatletter
\let\newfloat\newfloat@ltx
\makeatother
\usepackage{algorithm}
\usepackage{algpseudocode}
\usepackage{svg}
\usepackage{xcolor}
\usepackage{tabularx} 
\usepackage{booktabs} 
\usepackage{siunitx} 
\usepackage{xspace}
\usepackage{fontawesome}
\usepackage{hyperref}
\usepackage{cleveref}
\newcommand{\githubmaster}{\href{https://github.com/snehjp2/diff-stasis}{\faGithub}\xspace}

\makeatletter
\renewcommand{\thealgorithm}{\arabic{algorithm}}
\renewcommand{\fnum@algorithm}{\fname@algorithm~\thealgorithm:}
\makeatother

\newcommand{\gammaell}{\mathrm{\Gamma_\ell}}
\newcommand{\omegaell}{\mathrm{\Omega_\ell^{(0)}}}

\begin{document}

\preprint{APS/123-QED}

\title{On the Generality and Persistence of Cosmological Stasis}
\author{James Halverson}
\author{Sneh Pandya}%
\affiliation{%
NSF AI Institute for Artificial Intelligence and Fundamental Interactions (IAIFI) \\
Department of Physics, Northeastern University, Boston, MA 02115, USA 
}%

\begin{abstract}
Hierarchical decays of $N$ matter species to radiation may balance against Hubble expansion to yield stasis, a new phase of cosmological evolution with constant matter and radiation abundances. We analyze stasis with various machine learning techniques on the full $2N$-dimensional space of decay rates and abundances, which serve as inputs to the system of Boltzmann equations that governs the dynamics. We construct a differentiable Boltzmann solver to maximize the number of stasis $e$-folds $\mathcal{N}$. High-stasis configurations obtained by gradient ascent motivate log-uniform distributions on rates and abundances to accompany power-law distributions of previous works. We demonstrate that random configurations drawn from these families of distributions regularly exhibit many $e$-folds of stasis. We additionally use them as priors in a Bayesian analysis conditioned on stasis, using stochastic variational inference with normalizing flows to model the posterior. All three numerical analyses demonstrate the generality of stasis and point to a new model in which the rates and abundances are exponential in the species index. We show that the exponential model solves the exact stasis equations, is an attractor, and satisfies $\mathcal{N}\propto N$, exhibiting inflation-level $e$-folding with a relatively low number of species. This is contrasted with the $\mathcal{N}\propto \log(N)$ scaling of power-law models. Finally, we discuss implications for the emergent string conjecture and string axiverse. 
\begin{center}
\texttt{\string{\href{mailto:j.halverson@northeastern.edu}{j.halverson},\href{mailto:pandya.sne@northeastern.edu}{pandya.sne}\string}@northeastern.edu}, \githubmaster
\end{center}
\end{abstract}

\maketitle


\section{\label{sec:level1}Introduction}

\vspace{-.5cm}
One of the hallmark characteristics of an expanding universe is the time-evolution of components that contribute to the overall energy density of the universe. It has recently been introduced \citep{Dienes_2022} that this assumption is not always true, and in fact the universe could experience extended phases of cosmological ``stasis'' in which the cosmological abundances of matter, radiation, and/or vacuum energy can remain steady over extended $e$-folds of cosmological evolution, facilitated by various physical mechanisms driving energy pumps which oppose the effects of cosmological expansion \cite{dienes2023stasis}. These phenomena can arise naturally from a variety of beyond the Standard Model (BSM) physical theories, for instance in those that predict towers of unstable states that inevitably decay.

There are a number of different flavors of stasis.
In the original formulation
\citep{Dienes_2022},  stasis is achieved when a tower of $\phi_\ell$ matter states dominate the energy density but then hierarchically decays into radiation (hereby referred to as $M \rightarrow \gamma$ stasis), whereas \citep{dienes2023stasis} further introduced mechanisms of vacuum energy to matter ($\Lambda \rightarrow M$) and vacuum energy to radiation ($\Lambda \rightarrow \gamma$) stasis in a similar context, as well as studies of the dynamics of ``triple stasis'' with simultaneous $\Lambda \rightarrow M \rightarrow \gamma$ stasis. 
Such alternative cosmological histories can have observational consequences \cite{dienes2023primordialblackholesplace, dienes2024cosmologicalstasisdynamicalscalars} depending crucially on the flavor, timescale, and duration of stasis. Though in many cases these provide interesting possibilities for the evolution of our universe, stasis may also constrain ultraviolet-complete theories such as string theory if it arises after nucleosynthesis. We will comment on stasis in the context of Kaluza-Klein towers, string towers, and the string axiverse.

In this paper, we present an analysis of stasis to complement \citep{Dienes_2022}, focusing on $M\rightarrow \gamma$ stasis. This flavor of stasis depends crucially on the abundances $\Omega_{\ell}^{(0)}$ and decay rates $\Gamma_{\ell}$ of the $N$ particle species that matter dominate prior to stasis. Whereas \citep{Dienes_2022} laid out a general theory of strict stasis and derived many analytic results in an eight-dimensional power-law model, including attractor behavior, we focus on understanding stasis in a model-agnostic manner on the full $2N$-dimensional space of rates and abundances. Such an analysis seems to require numerics, for which we employ a variety of machine learning  tools (to optimize stasis $e$-folds, not cosmological viability; see Section \ref{sec:conclusions}), but lets us probe the generality of stasis and understand aspects of its duration.  The numerical methods uncover a new exponential model of stasis, which leads to extended periods of stasis.

This paper is organized as follows. First, we review essentials of $M\to \gamma$ stasis from \citep{Dienes_2022}. In Section \ref{sec:diffsim} we develop a differentiable Boltzmann solver that facilitates a number of numerical analyses involving stasis. Performing gradient ascent to maximize stasis, we see the emergence of exponential models that motivate log-uniform statistics. In Section \ref{sec:randomstasis} we study random stasis where rates and abundances are drawn from both these log-uniform distributions and also power-law distributions. Stasis occurs quite generally, with longer duration in the log-uniform case. In Section \ref{sec:svi} we use both types of distributions as priors for stasis-conditioned posteriors. The posteriors in this Bayesian analysis are modeled using a neural network known as a normalizing flow, which are optimized using stochastic variational inference. Posterior samples lead to more robust stasis, referring to longer a duration of stasis epochs, and again prefer an exponential model. In Section \ref{sec:models} we study the exponential model directly, demonstrating that it leads to parametrically-in-$N$ longer periods of stasis than power-law models, and discuss potential interfaces with string theory.

\subsection{\label{sec:level2}Matter-Radiation Stasis}

For a tower of states $\phi_{\ell}$ where $\ell \in [0,1,2,..., N]$, let $\rho_{\ell}$ denote a corresponding energy density and $\Omega_{\ell}$ a corresponding abundance. For general types of stasis, $\phi_{\ell}$ can either be a tower of massive scalar fields contributing to the total matter abundance $\Omega_{M}$ of the universe, or a tower of vacuum energy fields contributing to $\Omega_{\Lambda}$. Below, we will focus on the $M \rightarrow \gamma$ stasis formalism. Recall that for any energy density $\rho_i$, $\Omega_{i}$ is related via
\begin{equation}
\label{eqn:1}
    \Omega_i \equiv \frac{8 \pi G}{3 H^2} \rho_{i} \; ,
\end{equation}
where $H$ is the Hubble parameter and $G$ is Newton's gravitational constant. 

Differentiating, the time evolution of $\Omega_{i}$ is then
\begin{equation}
\label{eqn:2}
    \frac{d\Omega_i}{dt} = \frac{8 \pi G}{3} \left( \frac{1}{H^2} \frac{d \rho_{i}}{dt} - 2 \frac{\rho_i}{H^3} \frac{dH}{dt} \right) \; .
\end{equation} This set of Boltzmann equations is dependent on the time-evolution of individual $\rho_i$ as well as $H$. Using the Friedmann equation for $dH/dt$ in a Friedmann-Robertson-Walker (FRW) universe, we obtain
\begin{equation}
\label{eqn:3}
    \frac{dH}{dt} = -H^2 - \frac{4 \pi G}{3} \left( \sum_\ell \rho_\ell + 3 \sum_\ell p_\ell \right) \; ,
\end{equation}
where in practice this is expressed in terms of the equation-of-state parameter $w \equiv p_\ell / \rho_\ell$ for a component. Simplifying further, we arrive at
\begin{equation}
\label{eqn:finalhubble}
    \frac{dH}{dt} = -\frac{1}{2} H^2 (4 - \Omega_M)\;,
\end{equation} 
having invoked $w_\gamma = 1/3$ and $w_M = 0$ in equation \ref{eqn:3}. Integrating both sides, we arrive at
\begin{equation}
    H(t) = \frac{2}{4 - \langle \Omega_M \rangle} \left( \frac{1}{t - t^{(0)}}       \right) \;,
\end{equation}
where we have used the approximation that $H^{(0)} (t - t^{(0)}) \gg 1$ and $\langle \Omega_M \rangle$ is the time-averaged matter abundance defined as
\begin{equation}
   \langle \Omega_M \rangle =  \frac{1}{t - t^{(0)}} \int_{t^{(0)}}^t dt' \Omega_M (t') \;.
\end{equation} During stasis, $d\langle \Omega_M \rangle / dt = 0$, in which $\langle \Omega_M \rangle = \overline{\Omega}_M$, the asymptotic stasis abundance.

In $M \rightarrow \gamma$ stasis, the decays of individual matter species are what source the production of radiation in this universe. Stasis epochs necessarily require time-dependent energy densities, whose equations of motion in this case are given by
\begin{equation}
\label{eqn:rhoM}
        \frac{d\rho_\ell}{dt } = -3 H \rho_\ell - \Gamma_\ell \rho_\ell \;.
\end{equation} 
Returning to equation \ref{eqn:2} and substituting in the result of equation \ref{eqn:finalhubble} and \ref{eqn:rhoM}, we arrive at the set of N ODEs which govern the time evolution of individual $\Omega_\ell$
\begin{eqnarray}
\label{eqn:finalODE}
    \frac{d \Omega_\ell}{dt} = H \Omega_\ell \left( 1 - \Omega_M \right) - \Gamma_\ell \Omega_\ell \;.
\end{eqnarray}
Equation \ref{eqn:finalODE}, in combination with the ODE for the Hubble parameter directly gives the dynamics for our system. As our Universe only contains matter and radiation, $\Omega_M + \Omega_{\gamma} = 1$ at all times and the dynamics for radiation are easily obtained by recognizing that $d\Omega_M/dt = -d\Omega_{\gamma}/dt$. 

For $M \rightarrow \gamma$ stasis, the universe is beginning in a matter dominated state, $\Omega_M (t^{(0)}) = 1$. With time-evolution, the individual matter species gradually redshift or begin decaying into radiation; the effects of both must balance to have $d\Omega_M/dt$ = 0 during stasis. This will happen for all times when the decays of $\phi_\ell$ are exactly counterbalanced by Hubble expansion. Further, all individual $\Omega_\ell$'s during stasis must cooperate to produce an asymptotic abundance $\overline{\Omega}_{M}$. These combined form the two necessary and sufficient conditions for an extended period of stasis:
\begin{equation}
\label{eqn:stasiscondition}
    \sum_{\ell} \Gamma_\ell \Omega_\ell = H (\Omega_M - \Omega_M^2)
\end{equation}
\vspace{-1.4em}
\begin{equation}
\label{eqn:stasiscondition1}
    \sum_{\ell} \Omega_\ell (t) = \overline{\Omega}_{M} \;.
\end{equation}
Equation \ref{eqn:stasiscondition} as written is actually a condition for \emph{eternal} stasis, which of course cannot be physical for some finite tower of states. However, we can illuminate how the stasis epoch ends by operating under the assumption of eternal stasis. Let us assume we're in a period of stasis where $\Omega_M$ has achieved its $\overline{\Omega}_{M}$ stasis value. With these conditions, we can study the explicit time dependence for many of the quantities of interest. Beginning with the Hubble parameter in equation \ref{eqn:3}, the solution is
\begin{equation}
\label{eqn:hubble}
    H(t) = \left( \frac{2}{4 - \overline{\Omega}_{M} } \right) \frac{1}{t} \;,
\end{equation}
which further implies that the scale factor grows as
\begin{equation}
\label{eqn:scalefactor}
    a(t) = a_* \left( \frac{t }{t_*} \right)^{2 / (4- \overline{\Omega}_M)}
\end{equation}
for a fiducial time $t_*$. It further follows from equation \ref{eqn:rhoM} that
\begin{equation}
    \rho_\ell(t) = \rho_\ell^* \left( \frac{t}{t_*}\right)^{-6 / (4- \overline{\Omega}_M)} e^{-\Gamma_\ell (t - t_*)} \; ,
\end{equation}
which in turn implies that
\begin{equation}
\label{eqn:fiducialomega}
    \Omega_\ell(t) = \Omega_\ell^*\left( \frac{t}{t_*} \right)^{2 - 6/(4-\overline{\Omega}_M)} e^{-\gammaell(t-t_*)} \; .
\end{equation}
The result of equation \ref{eqn:hubble} when inserted into equation \ref{eqn:stasiscondition} while assuming a period of stasis gives
\begin{equation}
\label{eqn:stasiscondition2}
\sum_{\ell} \Gamma_{\ell} \Omega_{\ell} = \frac{2 \overline{\Omega}_M (1 - \overline{\Omega}_M)}{4 - \overline{\Omega}_M} \frac{1}{t} \; ,
\end{equation} exhibiting a power-law dependence for $t$, which cannot be true for all $t$. Thus, this will not yield an eternal stasis epoch, but a stasis epoch which is terminated when all species decays have concluded.

The models studied in \citep{Dienes_2022, dienes2023stasis} consider a spectrum of decay widths \{$\Gamma_\ell$\} and abundances \{$\Omega_\ell$\} motivated by a variety of BSM models which follow a power-law scaling
\begin{equation}
    \Gamma_\ell =  \Gamma_0 \left( \frac{m_\ell}{m_0}\right)^{\gamma}, \quad \Omega_\ell^{(0)} = \Omega_0^{(0)} \left(\frac{m_\ell}{m_0}\right)^\alpha
\end{equation}
where the mass spectrum takes the form
\begin{equation}
\label{eqn:pwrlawmass}
    m_\ell = m_0 + (\Delta m)\ell^\delta
\end{equation}
and $\Omega_0^{(0)}$ is a normalization factor enforcing $\Omega_M(t^{(0)}) = 1$. The parameters $\alpha$, $\gamma$, and $\delta$ are further restricted to the following range:
\begin{equation}
    -\frac{1}{\delta} < \alpha \leq \frac{\gamma}{2} - \frac{1}{\delta} \;.
\end{equation}
This therefore defines a $8D$ model parameterized by 
\begin{equation}
    \{\alpha, \gamma, \delta, m_0, \Delta m, \Gamma_0, \Omega_0^{(0)}, t^{(0)}   \}
\end{equation}
which is crucially a subset of the full $2N$-dimensional input parameter space that we seek to study.

It is important to note that missing from this list of parameters if the initial value of the Hubble constant, $H^{(0)}$. This model of stasis has exhibited \emph{global attractor} properties that were extensively studied in \citep{Dienes_2022}. The initial timescale for $\phi_\ell$ decays is dictated by the ratio $\Gamma_{N-1} / H^{(0)}$. When this ratio is small, the decays begin slowly after the starting time $t^{(0)}$, and the so-called ``edge effects'' in \citep{Dienes_2022} are mild. Conversely, when $\Gamma_{N-1} / H^{(0)} \gg 1$, particle decays begin almost immediately and there are severe edge effects. These edge effects are indeed necessary for a stasis state to both begin and end, as the condition in equation \ref{eqn:stasiscondition2} cannot be true when the decay process is just beginning or has concluded. However, due to the global attractor nature of stasis it is possible to achieve the same configuration of stasis, with the exception that the stasis state is approached from \emph{below} rather than above.

\section{\label{sec:diffsim} Maximizing Stasis with Differentiable Simulations}

The duration of stasis may be determined by solving the $N+1$ Boltzmann equations on a $(2N+1)$-dimensional parameter space of decay rates, abundances, and the initial value of the Hubble parameter. This problem is difficult due to its high dimensional nature, and in general we would also like to be able to differentiate the numerical solution to the Boltzmann equations to understand how the duration of stasis responds to variations in the rates and abundances.  A numerical Boltzmann solver may be thought of as a type of simulator, and we seek to differentiate through the entire simulation process.

Differentiable simulators are part of a growing trend in ML applications within the sciences, motivated by the emergence of more powerful and robust simulations. They have accelerated scientific analyses from molecular dynamics \citep{jaxmd2020} and biology to physics and cosmology \citep{Campagne_2023}. Further, with the advent of neural-network based probabilistic modeling, differentiable simulations are essential for implementing techniques such as stochastic variational inference \citep{hoffman2013stochasticvariationalinference} and Hamiltonian Monte Carlo \citep{Duane:1987de}. Differentiable simulations thus serve as powerful scientific tools for which to do both purely data-driven and probabilistic modeling studies. 

For $N$ species, our task is to solve a set of $N+1$ coupled ODEs as given in equations \ref{eqn:finalhubble} and \ref{eqn:finalODE} and compute the number of $e$-folds of a potential stasis epoch. We utilize \texttt{diffrax} \citep{kidger2021on}, a \texttt{jax}-based \citep{jax2018github} library that provides numerical differential equation solvers that preserve gradients and also allow backpropagation through our solutions to the Boltzmann equations. \texttt{jax} is a differentiable high-performance numerical computing library which utilizes just-in-time (\texttt{JIT}) compilation and vectorized computations, earning popularity in the sciences for its efficiency and modularity. The set of ODEs is stiff, meaning certain numerical techniques will be unstable without a sufficiently small resolution for the step-size. 
We use the \texttt{Kvaerno5} \citep{kvaerno2004singly} solver which is a 5th order explicit singly diagonal implicit Runge-Kutta (ESDIRK) method suited for stiff ODEs, with an absolute and relative tolerance $\text{atol} = \text{rtol} = 10^{-8}$ in the step size. We terminate the solver at $t = t_{max}$ when $\Omega_{M}(t) = 10^{-4}$, indicating that the universe has passed into a radiation-dominated epoch. Backpropagation through a differential equation can also be very memory expensive. We use \texttt{RecursiveCheckpointAdjoint} \citep{stumm2010new, wang2009minimal} which utilizes a binomial checkpointing scheme to preserve memory usage during backpropagation.

We seek to utilize the gradients in our solver to guide the parameter space towards stasis configurations. It is therefore not enough to solve the Boltzmann equations and be able to backpropagate through them, we also define a \emph{differentiable} algorithm to compute the stasis duration and asymptotic abundance from a given $\Omega_M$ curve. To this extent, we must first define a numerical notion of numerical stasis that can be applied to numerical $\Omega_M$ curves.

\subsection{\texorpdfstring{$\epsilon$-Stasis}{epsilon-Stasis}}

Stasis as defined up to this point is a strict condition on the time-evolution of a cosmological component. In \cite{Dienes_2022}, the model introduced is constructed so that this can be achieved exactly. However, from a numerical perspective, configurations that yield small deviations from exact stasis can nevertheless yield significant alterations to a cosmology. A more general definition of stasis is an epoch in which $\Omega_M$ is ``flat'' enough for ``long'' enough. This requires a parametric notion of stasis, and we therefore introduce an $\epsilon$-tolerance on stasis according to two different definitions. First, we develop a notion of stasis that allows us to create a differentiable (to enable the differentiable simulator) and accurate \emph{stasis finder} algorithm to isolate epochs of stasis. Then we present a more intuitive notion of stasis that is utilized in all the presentations of our results.

We begin with the notion of stasis that admits a differentiable stasis finder algorithm. Recall that stasis is a phenomenon induced by the cooperative behavior of individual $\Omega_\ell$. A period of stasis is therefore computed by analyzing the total abundance $\Omega_{M}(t) = \sum_\ell \Omega_\ell(t) $. For a given $\Omega_{M}(t)$ curve, we must compute the asymptote around which stasis occurs, and further isolate the duration of a stasis epoch. To do this in a differentiable way, we introduce an exponential weighting to compute a ``flatness score'' that yields the stasis duration in $t$
\begin{align}
\label{eqn:diffstasis}
&t_{s}(\Omega_{M}, \overline{\Omega}_{M}) = \left[ \sum_{i} \exp \left( -\frac{1}{\sigma} |\Omega_{M, i+1} - \Omega_{M, i}| \right. \right. \nonumber \\
&\quad \left. \left. - \frac{1}{\delta} |\Omega_{M, i} - \overline{\Omega}_{M}| \right) \right] \times \Theta((0.99 - \overline{\Omega}_{M})(\overline{\Omega}_{M} - 0.01)) \; ,
\end{align}
where $\Theta$ is the Heaviside function, enforcing that only mixed-component cosmologies are considered, and the index $i$ is that of the $i^\text{th}$ time-step in the solution $\Omega_M(t)$. A tolerance $\sigma = 0.02$ and $\delta = 0.09$, which roughly corresponds to a window-tolerance of $\pm 0.1$ about the asymptotic stasis abundance, was found to work best for optimization.

Upon termination of the solver, we use the solution for $H(t)$ to calculate the total number of $e$-folds of the simulation via
\begin{equation}
    \mathcal{N}_{max} = \int_{t^{(0)}}^{t_{\text{max}}} H(t) \, dt \;, 
\end{equation}
where it is assumed that $a(t^{(0)}) = 1$. This integral is evaluated numerically using the composite trapezoidal rule (i.e. \texttt{jax.numpy.trapezoid}).
It is then straightforward to compute the number of $e$-folds of stasis $\mathcal{N}$ via normalization with respect to the total number of $e$-folds, 
\begin{equation}
    \mathcal{N} = \mathcal{N}_{max} \cdot \frac{t_{s}}{t_{max}} \; .
\end{equation}
This normalization step allows us to isolate the contribution of stasis epoch relative to the entire simulation duration.

Intuitively, the stasis finder measures flatness by considering both how similar consecutive in $\Omega_{M,i}$ values are and how close these values are to $\overline{\Omega}_{M}$. It does so by creating a score that rewards sequences where values are close to each other (indicating flatness) and close to the target value (indicating relevance), adjusted by the parameters $\epsilon$ and $\delta$ to fine-tune sensitivity and scaling. 

This differentiable calculation, however, requires that $\overline{\Omega}_{M}$ is known. This is retrieved from a separate, differentiable abundance-finder algorithm. The algorithm first marks $\Omega_M$ values that are not matter dominated or radiation dominated (i.e. $0.01 < \Omega_M < 0.99$). It then determines links between consecutive valid values that are within a specified tolerance $\epsilon$. It proceeds to count these links for each time step and identifies the range of indices with the maximum number of links, indicating a period of stasis. Finally, it filters the valid $\Omega_{\text{M}}$ values within this range and computes $\overline{\Omega}_M$ as the median of these values, representing the typical abundance during the stasis period. This algorithm, of course, also yields a prediction for $\mathcal{N}$, but we find  by examining solutions that it is inaccurate and does not function as well as equation \ref{eqn:diffstasis}.

A pitfall of the differentiable stasis-finder in equation \ref{eqn:diffstasis} is that the sum of flatness scores considers \emph{all} parts of $\Omega_M(t)$ that are near the asymptotic abundance, as opposed to those \emph{only} within the stasis period. As such, the differentiable finder can be biased in $e$-folds; however, we will see that it still properly serves the purpose of a guiding the parameter space towards stasis configurations.

For these reasons, a more accurate, non-differentiable stasis finder which uses a sliding-window algorithm to isolate the longest stasis period is used in all presented results. This sliding window stasis finder is configured for a $10\%$-tolerance, e.g. the number of e-folds of stasis is determined by considering a window of $\pm 0.1$ around $\overline{\Omega}_M$ as valid. All the results presented in this paper use this notion of stasis.

\subsection{Maximizing Stasis}
\label{sec:stasiswithgrad}

We have amassed the necessary ingredients to use the differentiable simulation to optimize for stasis.
Using the simulation $\mathcal{S}(\theta = \{\Gamma_\ell, \Omega_\ell^{(0)}, H^{(0)}\})$, which returns $\mathcal{N}$ and $\overline{\Omega}_M$ after solving the Boltzmann equations and its gradients $\nabla_{\Gamma_\ell}$ and $\nabla_{\Omega_\ell^{(0)}}$, we can employ gradient ascent to optimize on stasis $e$-folds.

We begin with a single vector of samples of $\omegaell$ and $\gammaell$ for a given $N$, initialized according to draws from any appropriate distribution. The Boltzmann equation for $\Omega_\ell$ is inherently dimensionless; the only dimensionful parameters are $H^{(0)}$ and $\Gamma_\ell$. We will take all $\gammaell$ to be in units of Planck mass $M_p$ and no greater than this scale, e.g., $\max(\gammaell) \leq 1 \; M_p$. Similarly, we will restrict decay rates to be no smaller than those corresponding to the current age of the universe, $\min(\gammaell) \geq 10^{-62} \; M_p$.

We consider both power-law and standard uniform distributions for initializing $\gammaell$ and $\omegaell$. The power-law distribution is representative of the original model of stasis in the limit that $\Delta m / m_0 \gg 1$, while the standard uniform respects minimal physical constraints (i.e. positive-definiteness and $\max(\gammaell) \leq 1 \; M_p$) on the parameters. Recall that $\omegaell$ are normalized upon entering the simulation such that $\Omega_M(t^{(0)}) = 1$.  To generate samples from a power-law distribution with a relative scaling $\ell^\beta$, we draw from a Pareto distribution with a shape parameter $\alpha_p = 1 / \beta$. These samples are then inverted to represent draws from a power-law distribution. Henceforth, we denote power-law distribution samples as $X \sim \ell^\beta$, with the understanding that they are generated as the inverse of samples $X \sim \text{Pareto}(\alpha_p = 1 / \beta)$.

After sampling but before entering the simulation, we sort the $\gammaell$, which is without loss of generality since the $\ell$ subscript amounts to a species definition. However, we also often sort the $\omegaell$, so that increasingly large rates correspond to increasingly large abundances. This introduces a non-trivial physics-motivated correlation between the parameters, and we henceforth refer to such samples as ``sort-correlated." When we speak of identically and independently draw (i.i.d.) parameters, we mean prior to sort-correlation, unless otherwise stated.

Sort-correlation may be performed with any off-the-shelf sorting algorithm, including but not limited to a simple \texttt{jax.numpy.sort}. As $\gammaell$ and $\omegaell$ are updated with gradient ascent, we would like to continue to enforce that their spectra are sort-correlated. Doing so crucially requires that they are sorted smoothly and differentiably, to not interrupt the flow of gradients during optimization. For this reason, we implement a custom differentiable bitonic sorting algorithm inspired by \citep{diffsort}. This algorithm works by recursively dividing an array into smaller sub-arrays, sorting them, and then merging them using compare-and-swap operations, the latter of which is modified to be differentiable. This sorting technique is also used when doing SVI experiments. 

One necessary constraint to recognize is that simply optimizing on stasis $e$-folds will encourage matter-dominated cosmologies, as species' time spent decaying detracts from time spent redshifting, which is what contributes to the overall stasis duration. As such, a stasis optimization condition $f$ is designed to enforce a mixed component cosmology 
\begin{align}
\label{eq:gradcondition}
f(\theta) &= \mathcal{N}(\theta) - \alpha \left[(\overline{\Omega}_M(\theta) - l)^2 + (\overline{\Omega}_M(\theta) - u)^2 \right] \\
&\quad \times \left(1 - \Theta(\overline{\Omega}_M(\theta) - l) \Theta(u - \overline{\Omega}_M(\theta)) \right) \nonumber \; ,
\end{align}
where we recognize that $\mathcal{N}$ and $\overline{\Omega}_M$ are outputs of $\mathcal{S}(\theta)$. Above, $l = 0.2$ and $u = 0.8$, defining bounds on the allowed matter abundance, and $\alpha$ specifies the regularization strength depending on the experiment, ensuring the constraint has a meaningful effect on the optimization landscape. The Heaviside function indicates that the penalty is only applied when $\overline{\Omega}_M$ is outside the allowed abundance window. It is now clear to see why the $\epsilon$-stasis finder algorithm needs to be differentiable in both $e$-folds and matter abundances, as the gradients of $\overline{\Omega}_M$ are necessary to study optimization of stasis.

When performing gradient ascent on stasis, the sequential gradient updates for $\Gamma_\ell$ and $\Omega_\ell^{(0)}$ are computed as
\begin{align}
\label{eqn:gradupdate}
&\Omega_{\ell,i+1}^{(0)} = \Omega_{\ell,i}^{(0)} + \eta(t) \nabla_{\Omega_{\ell,i}^{(0)}} f(\theta) \\
&\Gamma_{\ell,i+1} = \Gamma_{\ell,i} + \eta(t) \nabla_{\Gamma_{\ell,i}} f(\theta)
\end{align}
for a given step $i$, where $\eta(t)$ is a (potentially) time-dependent learning rate. In subsequent experiments, a decay-factor $\gamma$ is applied at epoch $t'$ such that the learning rate has the functional form
\begin{equation}
\eta(t) = 
    \begin{cases}
    \eta_0, & \text{for } t < t' \\
    \gamma \cdot \eta_0, & \text{for } t \geq t'
\end{cases}
\end{equation}
Optimization is subject to an early-stopping criterion if $f(\theta)$ does not improve over a specified number of epochs $\xi$ or if a \texttt{NaN} is encountered during optimization. 

A detailed algorithm of optimizing stasis with gradient ascent is shown in Algorithm 1. In short, the algorithm 1) generates initial samples for rates and abundances; 2) sort-correlates them; 3) solves the Boltzmann equations; 4) uses the differentiable stasis-finder to compute the number of stasis e-folds; 5) compute gradients through the solution; 6) updates the parameters according to gradient ascent on stasis. This yields the parameters for the next iteration of the pipeline.

\begin{algorithm}
\caption{Gradient Ascent on Stasis}
\label{alg:algo1}
\begin{algorithmic}[1]
\Require $\theta = \{\gammaell, \omegaell \}$ simulation parameters, $\mathcal{N}$ stasis $e$-folds, $\alpha$ penalty coefficient, $l$ lower bound on matter abundance, $u$ upper bound on matter abundance, $\eta(t)$ learning rate, $\xi$ early-stopping threshold
\State Initialize: $\Gamma_\ell$, $\Omega_\ell^{(0)}$ via sampling and sort-correlate their spectra
\While{not converged}
    \State $\mathcal{N}, \overline{\Omega}_M \gets \mathcal{S}(\theta)$ 
    \State $f(\theta) \gets \mathcal{N}(\theta) - \alpha \left[(\overline{\Omega}_M(\theta) - l)^2 + (\overline{\Omega}_M(\theta) - u)^2 \right] \times \left(1 - \Theta(\overline{\Omega}_M(\theta) - l) \Theta(u - \overline{\Omega}_M(\theta))\right)$ 
    \State $\nabla_{\Omega_\ell^{(0)}} f(\theta), \nabla_{\Gamma_\ell} f(\theta)$ \Comment{Compute gradients}
    \State $\Omega_{\ell,i+1}^{(0)} \gets \Omega_{\ell,i}^{(0)} + \eta(t) \nabla_{\Omega_{\ell,i}^{(0)}} f(\theta)$ \Comment{Update abundances}
    \State $\Gamma_{\ell,i+1} \gets \Gamma_{\ell,i} + \eta(t) \nabla_{\Gamma_{\ell,i}} f(\theta)$ \Comment{Update decay rates}
    \State $\Omega_1^{(0)} \leq \Omega_2^{(0)} \leq \ldots \leq \Omega_N^{(0)}$ \Comment{Differentiably sort $\Omega_\ell^{(0)}$}
    \State $\Gamma_1 \leq \Gamma_2 \leq \ldots \leq \Gamma_N$ \Comment{Differentiably sort $\Gamma_\ell$}
    \State $\Gamma_\ell \gets \text{clip}(\Gamma_\ell, 0, 1)$ \Comment{Ensure $\Gamma_\ell$ physical}
    \If{$f(\theta)$ does not improve over $\xi$ epochs or \texttt{NaN} is encountered}
        \State \textbf{break} \Comment{Early-stopping criterion}
    \EndIf
\EndWhile
\end{algorithmic}
\end{algorithm}

\medskip
We study the outcome of gradient ascent optimization for initializations from a power law distribution corresponding to $\Gamma_\ell \sim \ell^3$ and $\Omega_\ell^{(0)} \sim \ell^1$ and a uniform initialization where $\Gamma_\ell, \Omega_\ell^{(0)} \sim \text{Uniform}(0,1)$. We conduct the experiments for $N=50$ species and with $\Gamma_{N-1} / H^{(0)} = 0.1$. We optimize both for a total of 50000 epochs with an early stopping threshold of $\xi = 2000$ epochs. An initial learning rate of $\eta = 0.01$ is used.

\begin{figure}
    \centering
    \includegraphics[width=\columnwidth]{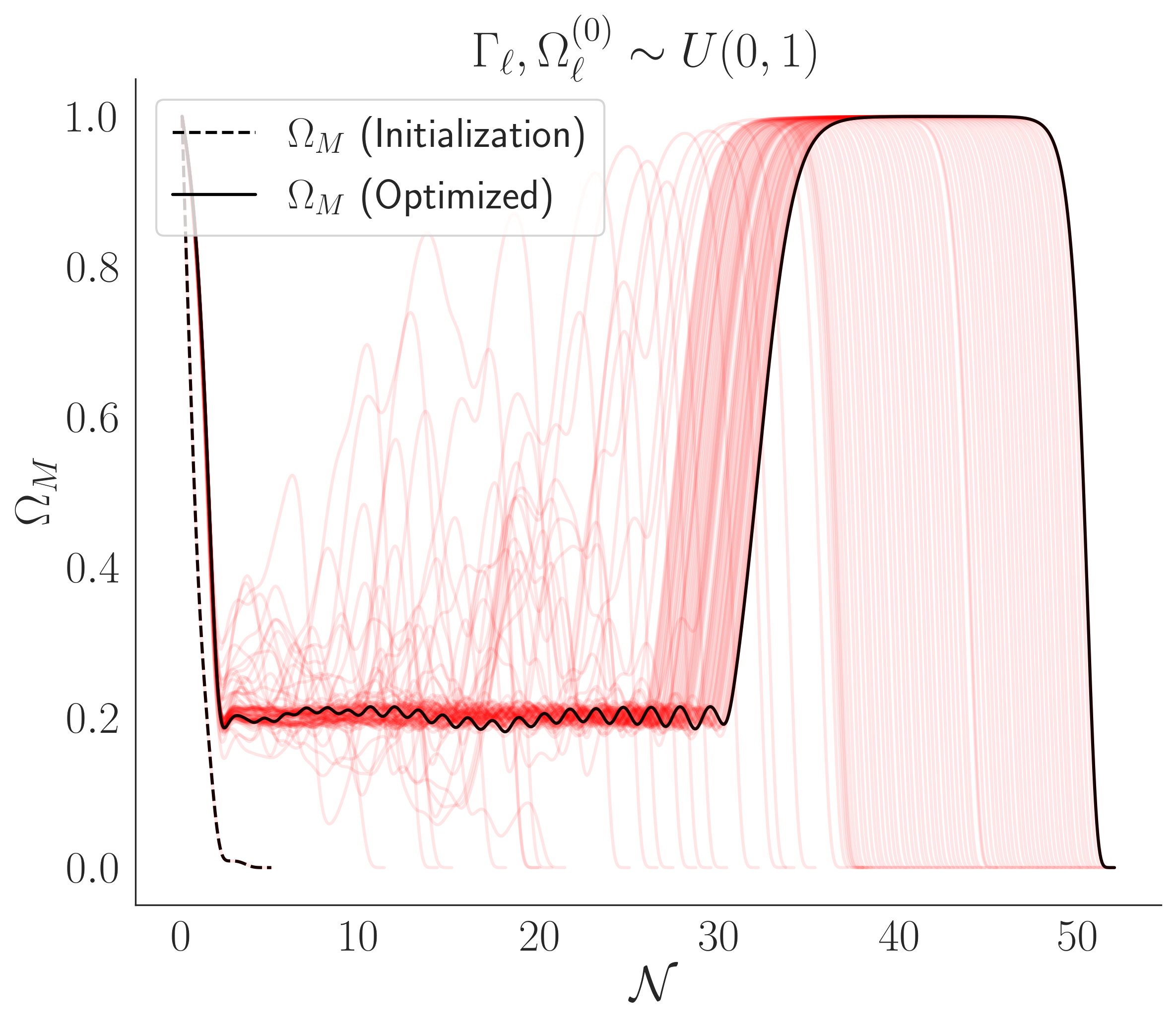}
    \caption{Gradient ascent trajectories for uniform initialization with $N=50$ species and $\Gamma_{N-1} / H^{(0)} = 0.1$. Parameters were optimized subject to the condition in equation \ref{eq:gradcondition} for 50,000 epochs of optimization with $\alpha = 10$. Intermediate trajectories for this initialization are relatively noisy, as seen by the red lines.}
    \label{fig:gradtrajuniform}
\end{figure}

Example gradient ascent trajectories for the uniform initialization is shown in Figure \ref{fig:gradtrajuniform}, where the optimization has settled on $\overline{\Omega}_M = 0.2$ and has achieved 27 $e$-folds of stasis. The change from initialized to optimized parameters is seen in going from the black-dashed line at initialization, through increasingly dark red intermediate trajectories as gradient ascent progresses, converging to the solid black line. From the trajectories, it can be seen that the optimization was relatively noisy. Indeed, from a numerical perspective what is essential for a robust epoch of stasis is large hierarchies in $\gammaell$ and smaller hierarchies in $\omegaell$, as is manifest for the power-law model of stasis and not characteristic of draws from a uniform distribution. Conceptually, this means that large changes are required to achieve stasis.

\begin{figure}
    \centering
    \includegraphics[width=\columnwidth]{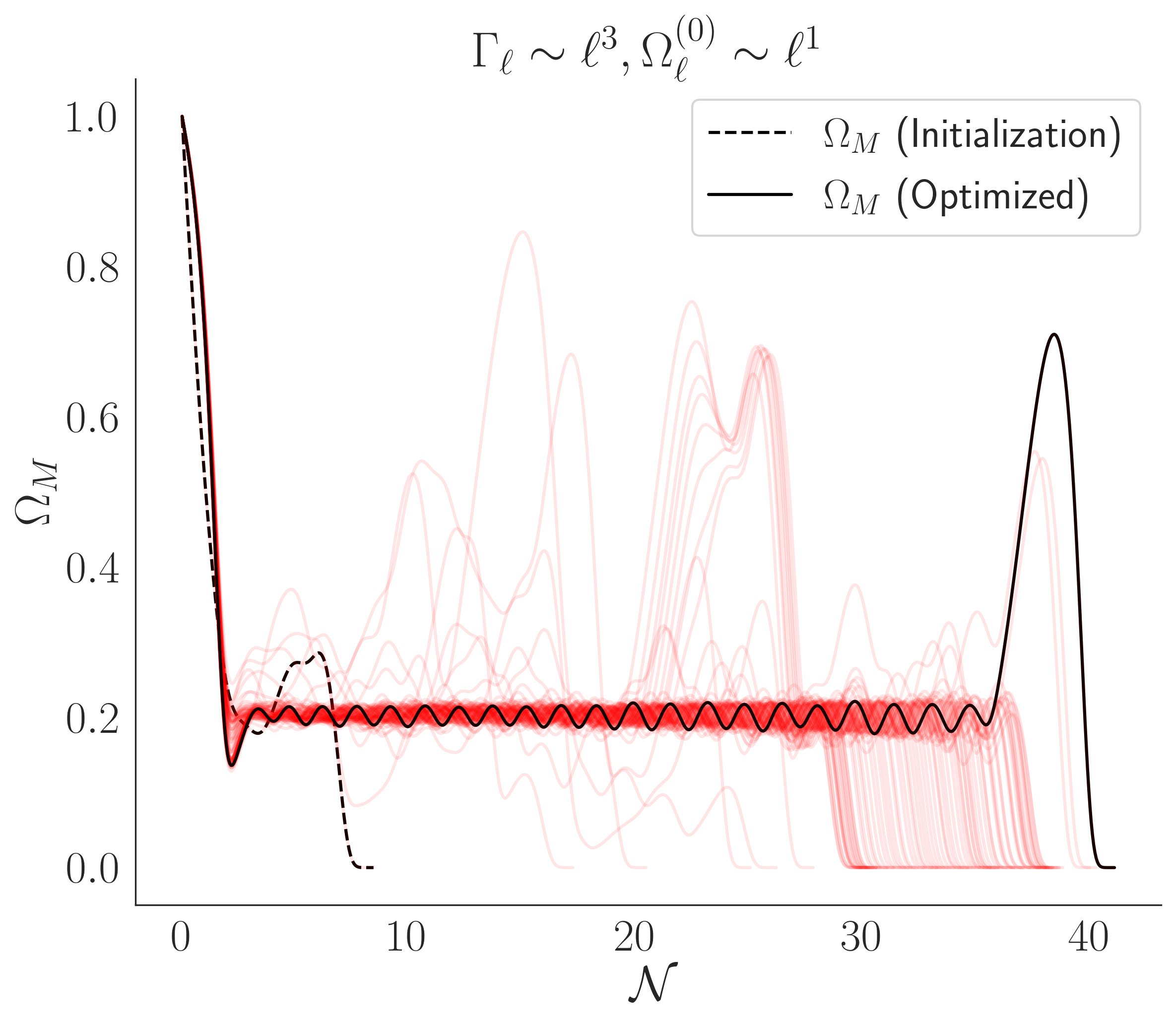}
    \caption{Gradient ascent trajectories for power-law initialization with $N=50$ and $\Gamma_{N-1} / H^{(0)} = 0.1$. 
    The benefit of decompressed $\Gamma_\ell$ and $\Omega_\ell^{(0)}$ spectra associated to power-law initialization is seen by the less noisy (red) background trajectories in the right panel and the more robust epoch of stasis when compared to the uniform initialization.}
    \label{fig:gradtrajpl}
\end{figure}

When initializing parameters as a power-law, the benefit of immediate hierarchies in the species spectra is apparent in Figure \ref{fig:gradtrajpl}. The optimization is more stable, as shown by the well-behaved intermediate values in red, and results in a longer epoch of stasis lasting 34 $e$-folds. It is then interesting to wonder: does a power-law initialization still result in a power-law model after following gradients, or does it change qualitatively? Similarly, does the uniform initialization change qualitatively under gradient ascent?

\begin{figure*}[!htbp]
    \centering
    \includegraphics[width=\textwidth]{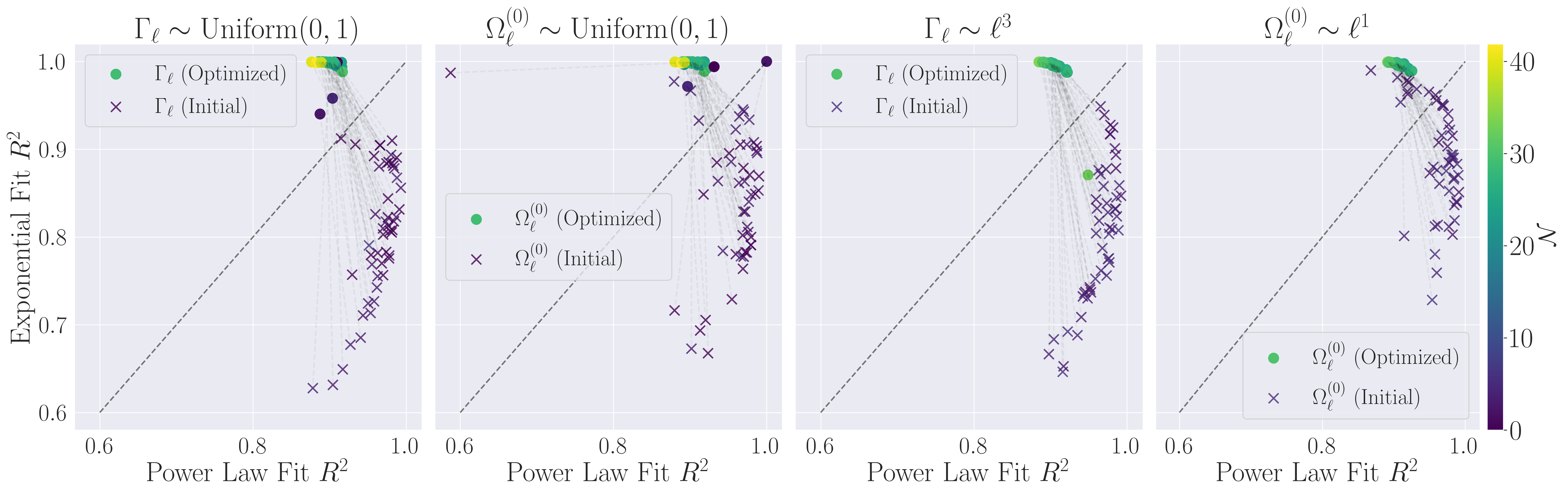}
    \caption{
    Experiments showing a preference for the exponential model upon optimizing stasis with gradients for uniform and power-law initializations for $N=50$ and $\Gamma_{N-1} / H^{(0)} = 0.1$. Gradient ascent was subject to the constraint $0.2 < \overline{\Omega}_M < 0.8$. Optimization was done for 50 random initializations for 50,000 epochs with early-stopping. An initial learning rate $\eta_0 =  0.01$ was used with a $\gamma = 0.1$ multiplicative decay at epoch $t' = 10,000$. A clear bias towards an exponential model is shown for optimal $\Gamma_\ell$ and $\Omega_\ell^{(0)}$, even when initialized with a power-law distribution similar to the original model of stasis. It is also seen that the drift shifts across the 1:1 line dividing exponential and power-law confidence equality. In some instances, the initialized parameters are shown to already be a good exponential fit due to the compression of the relative abundance and decay spectra. Even under such conditions, the drift towards a more exponential model and away from a power-law model is evident.}
    \label{fig:r2plot}
\end{figure*}

We can further deploy the differentiable simulation to study the model dependence of the optimized $\Gamma_\ell$ and $\Omega_\ell^{(0)}$. We optimize 50 random initializations of $\theta = \{\Gamma_\ell, \Omega_\ell^{(0)} \}$ with the differentiable simulation initialized according to a uniform distribution, $\Gamma_\ell, \Omega_\ell^{(0)} \sim U(0,1)$, and power-law distributions, $\Gamma_\ell \sim \ell^3$ and $\Omega_\ell^{(0)} \sim \ell^1$.  We conduct our experiments for $N=50$ and $\Gamma_{N-1} / H^{(0)} = 0.1$, the results of which are shown in Figure \ref{fig:r2plot}.

We can quantify the model dependence of $\gammaell$ and $\omegaell$ to a power-law model with linear fit in log-log space when looking at the functional dependence of $\gammaell$ and $\omegaell$ with $\ell$. Similarly, a linear fit in semi-log space is indicative of an exponential dependence with $\ell$. We use the coefficient of determination ($R^2$ score) for comparisons of model dependence across distributions and across scales, as it enjoys a scale invariance while still encoding information of fit residuals. The scale invariance is crucial, as the relative scale of optimized $\gammaell$ and $\omegaell$ are a priori different than from their respective initializations. Traditional metrics such as mean squared error are sensitive to this, and can yield misleading results.

In Figure \ref{fig:r2plot} we see that at initialization there is a clear preference towards a power-law model over exponential, as both $\Omega_\ell^{(0)}$ and $\Gamma_\ell$ generally exhibit $R^2 > 0.9$, while exponential fits feature $R^2 > 0.6$. This is expected, since neither of the initialization distributions were exponential. After optimization, there is a clear drift towards perfect exponential fit ($R^2= 1$) across all experiments, indicating a clear preference towards an exponential model of stasis when optimizing on stasis $e$-folds, even when being initialized with a power-law distribution that is reflective of the existing model of stasis. It is also seen in Figure \ref{fig:r2plot} that the optimized values are much more abundant in stasis $e$-folds than their respective initializations.

This result suggests a new model of stasis for which 
\begin{equation}
    \Omega_\ell^{(0)} \propto e^{\alpha \ell} \qquad
    \Gamma_\ell \propto e^{\gamma \ell},
\end{equation} an exponential model which is qualitatively different than the power-law model introduced in \citep{Dienes_2022}. It also opens the door for additional physical mechanisms that can result in particle spectra that can induce a stasis state. We will comment on those physics models in Section \ref{sec:models}.

We emphasize that this result has very little model bias: no trained neural networks or strong prior beliefs that restrict the effective parameter dimension were used in arriving at this result. The only assumption is a prior on the full parameter space from which the initial parameter are drawn, and then we simply differentiably simulate the stasis phenomenon conditioned on increasing stasis. On statistical grounds, these results suggest a new \emph{distribution} that these parameters follow; that is, a log-uniform distribution, a simple distribution for which samples are uniformly distributed in orders of magnitude. We will further study this model of stasis, and more generally the stasis parameter space from a distributional point of view, using neural networks in a Bayesian inference setting. To that end, these results suggest a new prior distribution.

\begin{figure*}[htbp!]
    \centering
    \includegraphics[width=\textwidth]{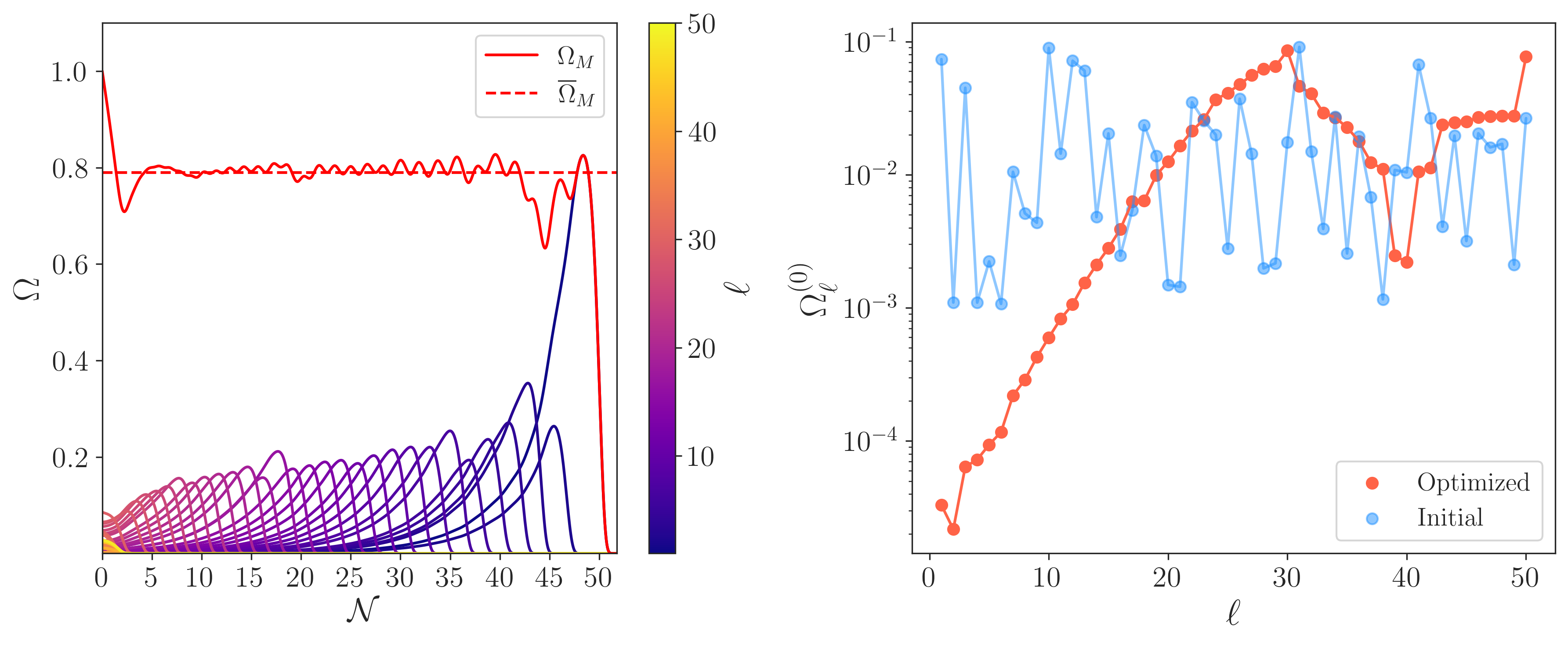}
    \caption{Example stasis epoch and gradient ascent trajectories for unsorted abundances with $N=50$ and $\Gamma_{N-1} / H^{(0)} = 0.1$. The initializations chosen were $\gammaell \sim \text{Log-U}(10^{-62}, 10^0)$ and $\omegaell \sim \text{Log-U}(10^{-2}, 10^0)$. $\gammaell$ samples were sorted before optimization via gradient ascent, which is without loss of generality as $\ell$ is a species definition. $\omegaell$ are left unsorted. We see that the gradients learn to correlate lower $\ell$ species which contribute to the overall stasis duration, resulting in a period of stasis lasting $\sim 43$ $e$-folds at an abundance of $\overline{\Omega}_M = 0.80$.  We see in the right panel that these contributing species approximately follow an exponential, characterized by a linear dependence on $\ell$ on a semi-log plot. For high $\ell$ species, which decay at early times to set the stasis abundance $\overline{\Omega}_M$, the algorithm does not learn to sort them. Indeed, $\sim 40\%$ of abundances do not monotonically increase with $\ell$.}
    \label{fig:unsortedomega}
\end{figure*}

\begin{figure*}[htbp!]
    \centering
    \includegraphics[width=\textwidth]{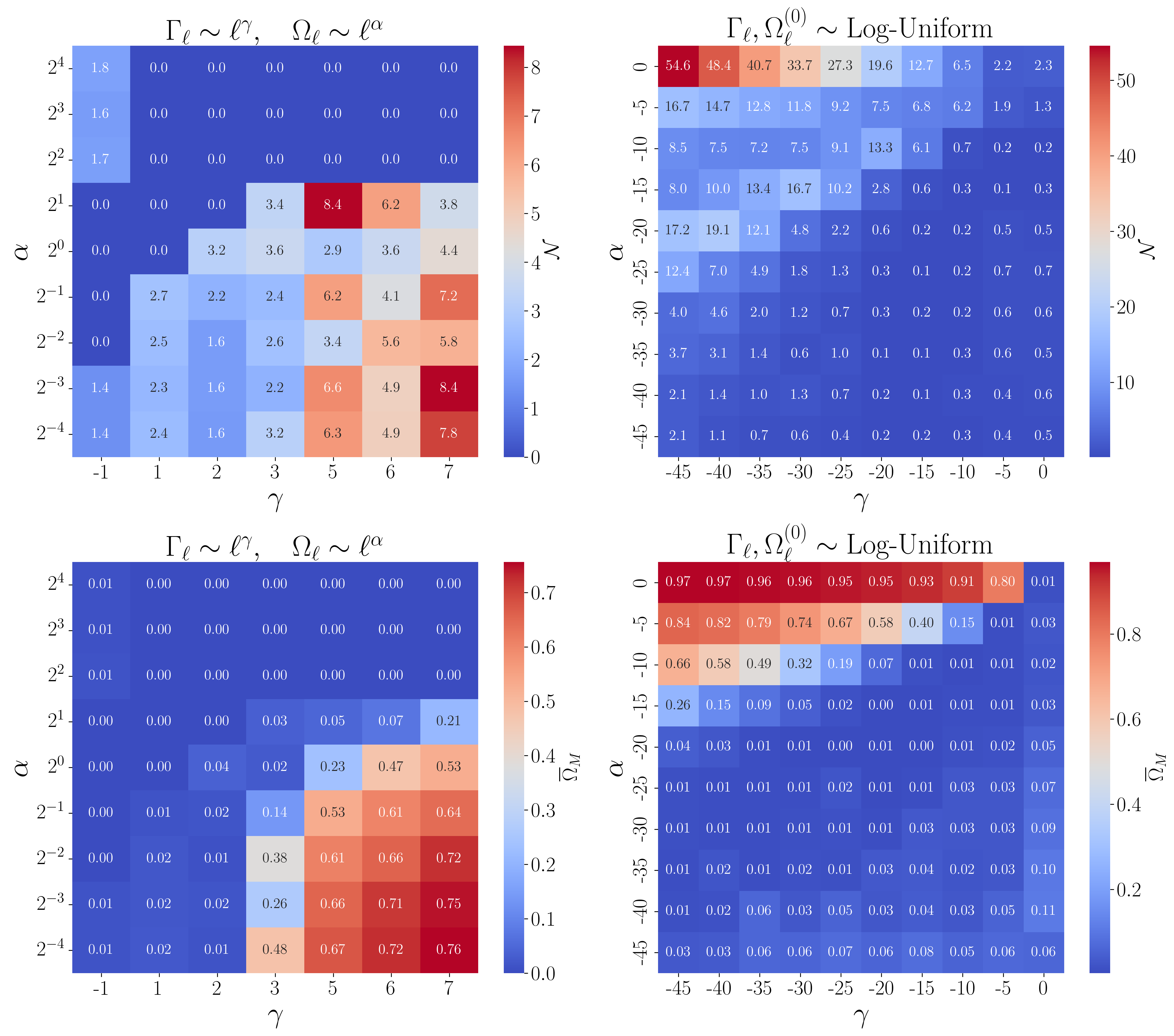}
    \caption{(\textbf{Left}) 
    Stasis configurations for $N=100$ species initialized with power-law prior draws across 100 realizations for $\omegaell$ and $\gammaell$, which correspond to the model of stasis introduced in \citep{Dienes_2022}. $\omegaell$ and $\gammaell$ are additionally sort-correlated before entering the simulation. The (non-differentiable) sliding-window stasis finder was used with a $10\%$-tolerance. We see that the maximum number of $e$-folds for this prior distribution is $\mathcal{N} \sim 8.4$ $e$-folds for a more matter-dominated cosmology. (\textbf{Right}) Stasis configurations for $N=100$ species initialized with exponential prior draws across 10 realizations, which correspond to sorted samples from a log-uniform distribution. The axes values correspond to $\omegaell \sim \text{Log-U}(10^{\alpha}, 10^0)$ and $\gammaell \sim \text{Log-U}(10^{\gamma}, 10^0)$, with the values chosen to illustrate the transition in $\mathcal{N}$ and $\overline{\Omega}_M$. The maximum number of $e$-folds for this prior is $\mathcal{N} \sim 55.1$ $e$-folds, a noticeably longer stasis duration than the power-law distribution and completely matter-dominated. Both distributions also feature a disallowed region in $\omegaell$, in which the abundance spectrum becomes sufficiently stressed that stasis is not possible. }
    \label{fig:heatmapcomparison}
\end{figure*}

\subsection{Stasis and Unsorted Abundances}

We have so far exclusively operated under the setting of sort-correlated $\gammaell$ and $\omegaell$. It is nonetheless interesting to study the emergence of a stasis epoch without respecting strict physical correlations between the parameters. As discussed, we may sort the $\gammaell$ without loss of generality, but in this section we keep $\Omega_\ell^{(0)}$ unsorted.

We proceed by initializing the decay rates as $\Gamma_\ell \sim \text{Log-Uniform}(10^{-62}, 10^{0})$ and sorting the spectrum, and $\Omega_\ell^{(0)} \sim \text{Log-Uniform}(10^{-2}, 10^0)$ for $N=50$ species, evolving the parameters according to the gradient update rules given in equations \ref{eqn:gradupdate} for 50,000 epochs with early-stopping. 

The result of this experiment is shown in Figure \ref{fig:unsortedomega}. This analysis has indeed shown that \emph{robust stasis epochs are possible without strictly correlated decay rates and abundances}, even though the species contributing to stasis have abundances increasing with decay rates. This realization would have been difficult to make with analytical modeling! 

We see in Figure \ref{fig:unsortedomega} that the randomly distributed abundances upon initialization become roughly monotonic up to $\ell \sim 30$, which are exactly the species in the left panel that balance to produce the stasis epoch lasting 43 $e$-folds. With the spectrum viewed on a semi-log plot in the right panel of Figure \ref{fig:unsortedomega}, it is clear to see that the contributing species follow an exponential model. The species $\ell > 30$ are those that decay at early times, setting the $\overline{\Omega}_M$ value. By just following gradients, it was deduced that these species did not in fact need to respect strict correlations with $\gammaell$ to produce stasis.

\section{Random Stasis in Physics-Motivated Distributions}
\label{sec:randomstasis}

The results of simply following  gradients in the differentiable simulation motivates an exponential model, as opposed to the original power-law model of \cite{Dienes_2022}. In a statistical language, this motivates the analysis of log-uniform priors in addition to the power-law priors. Power-law models were argued to be physically relevant in \cite{Dienes_2022}, and we will argue in Section \ref{sec:models} that exponential models are also physically motivated.

In this section, we study stasis in random draws from these physically-motivated priors, and argue that they are also statistically well-motivated in Bayesian statistics. First we will provide a statistical perspective on the priors, and then study stasis.

\subsection{Scale-Invariant Priors and Decay Rates}

A prior on a vector of random variables $\boldsymbol{\theta}$ is a density that represents some knowledge of the system or beliefs about the way it behaves, before further evidence is introduced.
There exists a particular class of priors known as ``uninformative priors,'' and a subclass of such priors that are scale-agnostic.  

One of these scale-invariant priors is known as the Jeffreys prior.
Mathematically, it is defined to exhibit the scaling 
\begin{equation}
      p(\boldsymbol{\theta}) \propto \sqrt{\det \mathcal{I}(\boldsymbol{\theta})}
\end{equation}
for its density function, where $\mathcal{I}(\boldsymbol{\theta})$ is the Fisher information metric defined as
\begin{equation}
    \mathcal{I}(\boldsymbol{\theta})_{ij} = \mathbb{E} \left[ \left( \frac{\partial}{\partial \theta_i} \log f(\boldsymbol{X}| \boldsymbol{\theta}) \right) \left( \frac{\partial}{\partial \theta_j} \log f(\boldsymbol{X}| \boldsymbol{\theta}) \right) \Bigg| \boldsymbol{\theta} \right]
\end{equation}
for a given likelihood function of parameters $f(\boldsymbol{X} | \boldsymbol{\theta})$, which is the probability density of a set of observed data $\boldsymbol{X}$ given parameters $\boldsymbol{\theta}$. We see that the Jeffreys prior is the volume measure with respect to the Fisher information metric. It is therefore diffeomorphism invariant and has the significant advantage that it does not depend on the choice of coordinates for model parameters.

We will now proceed to derive an appropriate Jeffreys prior for $\Gamma_\ell$. The time until decay $\tau$ for a particle given a decay rate $\Gamma$ is governed by the PDF
\begin{equation}
    f(\tau | \Gamma) = \Gamma \; e^{-\Gamma \tau} \;.
\end{equation}
Under conditions in which a likelihood function is twice differentiable and has a vanishing expectation value of the gradient of the log-likelihood, a simpler definition of the Fisher information can be used. By inspection, we see that $f(\tau | \Gamma)$ is twice differentiable with respect to $\Gamma$. It only remains to check that
\begin{align}
    \mathbb{E} \left[ \frac{\partial}{\partial \Gamma} \log f(\tau | \Gamma) \right] 
    &= \mathbb{E} \left[ \frac{1}{\Gamma} - \tau \right] \\
    &= \frac{1}{\Gamma} - \frac{1}{\Gamma} \nonumber \\
    &= 0 \; , \nonumber
\end{align}
where we have invoked that $\mathbb{E}[\tau] = 1 / \Gamma$, as $\tau$ follows an exponential distribution with respect to $\Gamma$. With these conditions satisfied, we can use a simpler form of the Fisher information
\begin{equation}
    \mathcal{I}(\boldsymbol{\theta})_{ij} = -\mathbb{E} \left[ \frac{\partial^2}{\partial \theta_i \partial \theta_j} \log f(\boldsymbol{X}; \boldsymbol{\theta}) \Bigg| \boldsymbol{\theta} \right]
\end{equation}
in deriving the Jeffreys prior for $\gammaell$.

The likelihood of observing a set of particles for a period of time $\boldsymbol{\tau} = \{\tau_1, \tau_2, ..., \tau_N\}$ before they decay with i.i.d decay rates $\boldsymbol{\Gamma} = \{\Gamma_1, \Gamma_2, ..., \Gamma_N\}$ is given by
\begin{equation}
    p(\boldsymbol{\tau} | \boldsymbol{\Gamma}) = \prod_{i=1}^{N} f(\tau_i | \Gamma_i)= \prod_{i=1}^{N} \Gamma_i \; e^{-\Gamma_i \tau_i} \; .
\end{equation}
Exploiting properties of logarithms, we arrive at
\begin{equation}
    \log p(\boldsymbol{\tau} | \boldsymbol{\Gamma}) = \sum_{i=1}^{N} \left(\log \Gamma_i - \Gamma_i \tau_i \right) \;,
\end{equation}
which allows one to compute the Fisher information matrix $I(\boldsymbol{\Gamma})_{ij}$ as
\begin{align}
    I(\boldsymbol{\Gamma})_{ij} &= -\delta_{ij}\mathbb{E}_{\boldsymbol{\tau}}\left[\frac{\partial^2 \log p(\boldsymbol{\tau} | \boldsymbol{\Gamma})}{\partial \Gamma_i \partial \Gamma_j} \right] \Rightarrow \\
   I(\boldsymbol{\Gamma})_{ii} & = -\mathbb{E}_{\boldsymbol{\tau}}\left[-\frac{1}{\Gamma_i^2}\right] = \frac{1}{\Gamma_i^2} \;,
\end{align}
where $\delta_{ij}$ is the Kronecker delta, enforcing that the $\Gamma_i$'s are i.i.d. distributed. Therefore, we see that a Jeffreys prior for $\Gamma_\ell$ must obey
\begin{equation}
\label{eqn:jeffreysgamma}
    p(\boldsymbol{\Gamma}) \propto \sqrt{\det I(\boldsymbol{\Gamma})} = \prod_{i=1}^N \frac{1}{\Gamma_i}.
\end{equation}
Such a property is exactly true for a log-uniform prior, as defined by its probability density function (PDF)
\begin{equation}
\label{eq:27}
f(x) = \frac{1}{{x \log\left(\frac{b}{a}\right)}}, \quad \text{for} \; a \leq x \leq b; \; a > 0
\end{equation}
where the interval $[a,b]$ is known as the \emph{support} and the term $\log(b/a)$ appears after normalizing \eqref{eqn:jeffreysgamma}.

It is remarkable that the distribution on $\Gamma_\ell$ motivated by flow on $\Gamma_\ell$ using the differentiable simulation aligns exactly with the Jeffreys prior for decay rates! We have thus established that such a distribution for decay rates is worthy of study, on both numerical and statistical grounds, with physics motivation in Section \ref{sec:models}.
 
\subsection{Random Stasis from Physical Priors}

It is now natural to study in detail how much stasis can emerge for $\Gamma_\ell$ and $\Omega_\ell^{(0)}$ spectra simply drawn from power-law and log-uniform priors. We study the mean $e$-folds of stasis that emerge across several configurations of the log-uniform and power-law prior, which are related to a new exponential model of stasis and the original model, respectively. 

We initialize the power-law priors according to draws $\Gamma_\ell \sim \ell^\gamma$ and $\Omega_\ell^{(0)} \sim \ell^\alpha$, where $\gamma \in [-1,7]$ and $\alpha = 2^k \; \text{for} \; k \in [-4,4]$. The procedure for generating power-law draws is given in \ref{sec:stasiswithgrad}. Similarly, the exponential model prior is initialized according to $\Gamma_\ell \sim \text{Log-Uniform}(10^{\gamma}, 10^0)$ and $\Omega_\ell^{(0)} \sim \text{Log-Uniform}(10^{\alpha}, 10^0)$ where $\gamma, \alpha = 5k \; \text{for} \; k \in \left[-9,0 \right]$. These samples are sort-correlated before entering the simulation. Experiments are run across 100 realizations for $N=100$ and with $\Gamma_{N-1} / H^{(0)} = 0.1$ while simultaneously enforcing $\max{(\Gamma_\ell)} = 1 \; M_p$.

We see in the top panel of Figure \ref{fig:heatmapcomparison} that the power-law prior achieves a maximum mean of $\mathcal{N} = 7.8$ $e$-folds of stasis for 
$\alpha = 2^{-3}$ and $\gamma = 7$. Indeed, we see that stretched spectra in $\Gamma_\ell$ and compressed spectra in $\Omega_\ell^{(0)}$ result in more matter-dominated cosmologies with longer epochs of stasis. Similarly, the exponential model prior achieves a maximum mean of $\mathcal{N} = 55$ $e$-folds for $\gamma = -45$ and $\alpha = 0$. It is important to note that after normalization, as $\Omega_\ell^{(0)}(t^{(0)}) = 1$, this configuration corresponds to a constant mass spectrum where $\Omega_\ell^{(0)} = 1/N$. This is indeed also a configuration in which the cosmology is matter dominated, as seen in the bottom panel of Figure \ref{fig:heatmapcomparison}. All configurations in Figure \ref{fig:heatmapcomparison} were computed using the sliding-window algorithm for computing $\epsilon$-Stasis, with $\epsilon = 0.1$.

The lower bounds of $\alpha$ and $\gamma$ in Figure \ref{fig:heatmapcomparison} are chosen to illustrate an effective cutoff in $\Omega_\ell^{(0)}$ in which the species completely decouple from the stasis phenomena. That is, their abundance spectra are so stretched that the individual species achieve their peak and decay such that any possible stasis epoch would occur with little to no matter in the Universe. In other words, in order for to achieve stasis with a non-trivial mixture of components, the prior on abundances should allow for an effective number of species  to be abundant enough at their peaks such that there is a possibility of a stasis epoch with $\overline{\Omega}_M > 0$. This is seen clearly in both choices of priors, with the cutoff occurring for $\alpha \leq -30$ and $\alpha \geq 4$. 

It is remarkable that stasis can emerge in some robust configurations with random (albeit sort-correlated) abundances and decay rates! We emphasize that there is no optimization or gradient information being used here; this is simply the result of sampling from distributions and sort-correlating the $\gammaell$ and $\omegaell$ spectra before entering the simulation. Further, it is interesting to see how persistent in $e$-folds an exponential model of stasis can be compared to a power-law; yielding close to 60 $e$-folds in certain cases with just $N=100$ species.

\section{\label{sec:svi}Stasis-Conditioned \\ Bayesian Posteriors}

We have so far operated in a data-driven setting without any explicit statistical modeling. In this section, we wish  to perform a Bayesian analysis on the complete $2N$-dimensional space of $\omegaell$ and $\gammaell$. Probabilistic inference on such a high-dimensional space using traditional techniques such as Markov Chain Monte Carlo would be computationally infeasible. Nonetheless, in the current age of machine learning, we can leverage techniques that exploit gradient information to make such tasks more accessible. To this end, we employ stochastic variational inference (SVI), a gradient-based inference technique that approximates complex probability distributions which are often intractable.

We would like to utilize SVI on the task of searching the full input parameter space of $\theta = \{\gammaell, \omegaell\}$ (keeping in mind that we fix the ratio $\Gamma_{N-1} / H^{(0)}$) by modeling the \emph{posterior} distribution $p(\theta |\mathcal{N} )$  conditioned on optimizing $\mathcal{N}$. At first glance, this seems like a tall order. The parameter space in question is $2N$-dimensional; it must be searched, optimized on, and result in an easily-sampled posterior. The crux of what makes this feasible is in the \emph{expressivity} of neural networks, combined with the \emph{information} of gradients. In a modern setting, SVI can transform Bayesian inference into a neural-network based optimization problem, thereby accelerating statistical modeling in high-dimensional spaces.

We begin with a review of the essentials of Bayesian and stochastic variational inference, including the evidence lower bound, and then apply these techniques in the context of stasis.

\subsection{The Evidence Lower Bound}
We proceed to define the fundamental object in SVI, which defines the objective function that is regressed on. All Bayesian inference problems begin with Bayes' Theorem, from which we have
\begin{equation}
\label{eqn:17}
    p(\theta | \mathcal{N}) = \frac{p(\mathcal{N} | \theta) p(\theta)}{p(\mathcal{N})} \; ,
\end{equation}
where $p(\mathcal{N} | \theta)$ is the likelihood and $p(\theta)$ is the prior distribution on $\theta$. In a traditional setting, where there are observations to guide the posterior towards, the likelihood encodes the probability of observed data given input parameters. It is sought to be maximized. For our purposes, the likelihood is computed via the simulation $\mathcal{S}(\theta)$ which outputs $\mathcal{N}$ $e$-folds of stasis. Since solving Boltzmann equations is deterministic, the simulation likelihood is exactly 
\begin{equation}
\label{eqn:llhood}
    p(\mathcal{N} |\Gamma_\ell, \Omega^{(0)}_\ell, H^{(0)}) =  \delta(\mathcal{N} - \mathcal{S}(\Gamma_\ell, \Omega^{(0)}_\ell, H^{(0)})) \; ,
\end{equation}
where $\delta$ is the Dirac-delta function. We will see that the simulation likelihood is disadvantageous as far as introducing information that aids in searching the full parameter space. For this reason, we adopt a simple optimization likelihood with the definition
\begin{equation}
\label{eqn:surllhood}
    p(\mathcal{N} |\Gamma_\ell, \Omega^{(0)}_\ell, H^{(0)}) \propto e^{\kappa \cdot  \mathcal{N}} \;,
\end{equation}
which can be potentially subject to a matter abundance constraint similar to that used in equation \ref{eq:gradcondition}. We emphasize the use of $\propto$ as this optimization likelihood is, of course, not a normalized probability density. Further, when implementing SVI, we find that the use of a numerical pre-factor $\kappa$ to increase the ``strength'' of the likelihood term is beneficial.  This surrogate (unnormalized) likelihood serves the utility of encouraging the posterior distribution to explore parameter space that yields stasis; the exact motivation behind this optimization likelihood will be expanded upon in section \ref{sec:theoryspaces}. The denominator of equation \ref{eqn:17}, $p(\mathcal{N}) = \int p(\mathcal{N} | \theta)p(\theta) d\theta$ is known as the \emph{Bayesian evidence} or \emph{marginal likelihood} and is in general computationally intractable, as it requires an integral over the entire parameter space. 

The crux of SVI is bypassing the direct application of Bayes' theorem by introducing a variational family $q_\phi(\theta | \mathcal{N})$ with variational parameters $\phi$, which should be distinguished from the parameters $\theta$ of the statistical model. The variational family can range from anything as simple as a standard Gaussian distribution with a trainable mean and variance parameter, to a complex neural network modeling a density on $\theta$ with millions of trainable parameters $\phi$. In this work, we will choose the latter.

We would like to maximize the similarity between the variational family and the true posterior by minimizing the Kullback-Leibler (KL) divergence between the two distributions, generally defined as
\begin{equation}
    D_{KL}(P||Q) = \int P(x) \log \left( \frac{P(x)}{Q(x)} \right) dx
\end{equation}
for two continuous distributions $P$ and $Q$. The KL divergence is a type of statistical ``distance", measuring how different the two distributions are. It is always positive semi-definite, satisfying $D_{KL}(P || Q) \geq 0$; however, it is not a metric in a formal sense as it is not a symmetric quantity (i.e. $D_{KL}(P || Q) \neq D_{KL}(Q || P))$. 

Proceeding with our derivation, after some algebra and substitutions of Bayes' Theorem we arrive at
\begin{align}
\label{eqn:19}
   D_{KL}(q_\phi(\theta) || p(\theta | \mathcal{N})) = & \ -\mathbb{E}_{q_\phi(\theta)} \left[\log p(\mathcal{N} | \theta) - \log q_\phi(\theta) \right. \nonumber \\
   & \left. + \log p(\theta) \right] + \log p(\mathcal{N}) \; ,
\end{align}
where the first term, known as the Evidence Lower Bound (ELBO), is the only term on that RHS that depends on the variational parameters. This is a fundamental object in SVI. It is easy to interpret the purpose of the ELBO when written as
\begin{align}
\label{eqn:ELBO}
    \text{ELBO}(\phi, \theta, \mathcal{N}) &= \mathbb{E}_{q_\phi(\theta)} \left[ \log p({\mathcal{N}} | \theta) - D_{KL}\left(q_\phi(\theta) \| p(\theta)\right) \right] \; ,
\end{align}
where the first term is the expected log-likelihood, introducing information from the likelihood (simulation) to the posterior, and the second term is the KL-divergence between the variational distribution and prior, which ensures the posterior distribution has knowledge of the prior distribution, as required by Bayes' theorem. We are then free to rewrite equation \eqref{eqn:19} as
\begin{align}
    D_{KL}(q_\phi(\theta)||p(\theta | \mathcal{N})) = -\text{ELBO}(\phi, \theta, \mathcal{N}) + \log p(\mathcal{N}) \; .
\end{align}
Thus, it is clear to see that minimizing the KL divergence is equivalent to maximizing the ELBO. Moreover, since the KL divergence is positive-definite, the ELBO is the lower bound of the evidence. If $q_\phi(\theta)$ approximates $p(\theta | \mathcal{N})$ well, computing the intractable evidence is no longer needed. 

We see that the ELBO is a function of both variational family parameters $\phi$ and simulation parameters $\theta$. It is a loss function that the variational family is regressing on, so its gradients must also be accessible. What must be computed to optimize $q_\phi(\theta)$ is the following:
\begin{equation}
\begin{aligned}
    &\nabla_{\phi} \text{ELBO}(\phi, \theta, \mathcal{N}) = \Bigg\{\underbrace{-\mathbb{E}_{q_\phi(\theta)} \left[ \nabla_\phi \log q_\phi (\theta) \right]}_{\text{Variational Distribution}}, \\
    & \hspace{0pt} 
    \underbrace{\mathbb{E}_{q_\phi(\theta)} \left[ \nabla_\phi \log p(\mathcal{N}|\theta) \right]}_{\text{Likelihood Term}} \Bigg\} \; ,
\end{aligned}
\end{equation}
where the expectation value is taken over $q_\phi(\theta)$. To backpropagate through the likelihood term in updating the variational family parameters, we use the chain rule
\begin{equation}
\nabla_\phi \log p(\mathcal{N}|\theta) = \nabla_\theta \log p(\mathcal{N}|\theta) \cdot \nabla_\phi \theta
\end{equation}
where we see the first term requires the gradients with respect to $\theta$:
\begin{equation}
\begin{aligned}
    &\nabla_{\theta} \log p(\mathcal{N}|\theta) = \Bigg\{
    \underbrace{\frac{\partial \log p(\mathcal{N}|\theta)}{\partial \Gamma_\ell}}_{\text{Decay Rates}},
    \underbrace{\frac{\partial \log p(\mathcal{N}|\theta)}{\partial \Omega^{(0)}_\ell}}_{\text{Abundances}} \Bigg\} \; .
\end{aligned}
\end{equation}
One can recognize from the optimization likelihood equation \ref{eqn:surllhood} that $\log p(\mathcal{N} | \theta) = \mathcal{N}$, which is precisely the output of the differentiable simulation $S(\theta)$. It is thus clear to see mathematically why gradients are necessary when implementing SVI --- in order to backpropagate through the likelihood to update $\phi$, the gradients with respect to $\theta$ must be known. 

\begin{figure*}[!htbp]
    \centering\includegraphics[width=\textwidth]{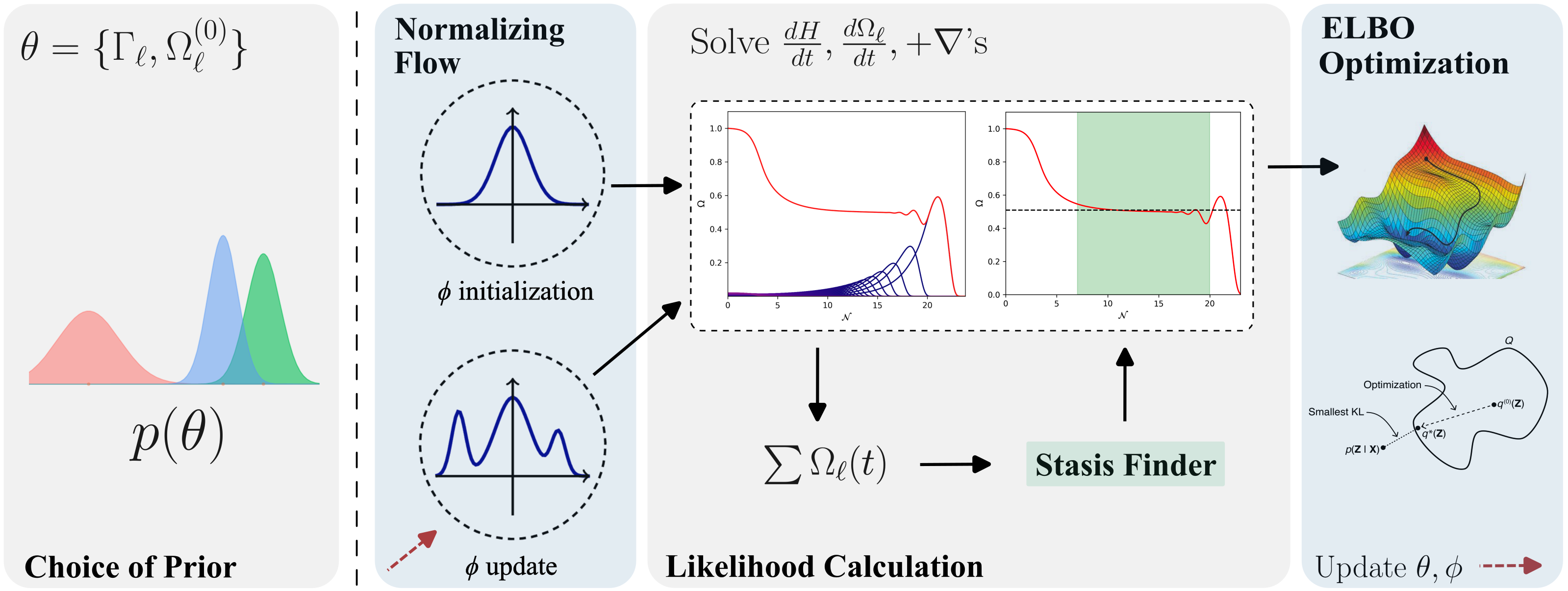}
    \caption{Stochastic variational inference pipeline. For a given experiment, a Bayesian prior in parameters is chosen which acts as a form of regularization in the ELBO. During training, parameters are sampled from the variational family $q_\phi(\theta)$, in this work chosen to be a Block Neural Autoregressive Flow (BNAF). Samples are differentiably sorted before entering the stasis simulation, which solves the set of $N+1$ coupled Boltzmann equations using \texttt{diffrax} to preserve the flow of gradients. The following $\Omega_M(t)$ curve is passed into the differentiable stasis finder to isolate the stasis $e$-folds $\mathcal{N}$ and the asymptotic matter abundance $\overline{\Omega}_M$. The stasis value is used in the likelihood calculation which is factored into the ELBO loss, which is used to iteratively optimize $q_\phi(\theta)$.}
    \label{fig:svipipeline}
\end{figure*}

\subsection{Normalizing Flows}
There are several options one can consider for constructing a variational family in SVI. One of the most expressive options is the normalizing flow (NF) \citep{rezende2016variationalinferencenormalizingflows}. This method operates by transforming samples $z$
from a simple probability density (e.g., Gaussian or Uniform) $q_0(z)$ into a complex posterior density $q_\phi(x)$. Here, $x$ represents the transformed variables in the target distribution, achieved through a series of invertible transformations. The invertibility is crucial because it allows for both forward and inverse mappings between the base distribution and the complex target distribution, generally enabling the calculation of probability densities. 

NFs are a class of bijective neural networks with trainable parameters $\phi$, where $q_{\phi}(x): \mathbb{R}^d \rightarrow \mathbb{R}^d$. To learn arbitrarily complex invertible functions, NFs are constructed as the composition of a series of $N$ invertible and bijective functions $f$:
\begin{equation}
    q_\phi(x) = f_N \circ f_{N-1} \circ \dots \circ f_1(z) \; ,
\end{equation}
leveraging the fact that the composition of a set of invertible functions is itself still invertible.

The mapping between two density functions $q_{\phi}(x)$ and $q_{0}(z)$ is related via the absolute value of Jacobian of the transformation
\begin{equation}
\label{eqn:flowtransfo}
    q_{\phi}(x) = q_{0}(z) \cdot \left| \det \left( \mathbf{J}_{q_{\phi}(z)} \right) \right| \;.
\end{equation}
Despite its simple functional form, the Jacobian computation can be prohibitively expensive for high-dimensional data or more complex architectures, where it is computed as
\begin{equation}
   \left| \det \left( \mathbf{J}_{q_{\phi}(z)} \right) \right| = \prod_{i=1}^N \left| \det \left( \frac{\partial f_i (z_{i-1})}{\partial z_i} \right) \right| \; .
\end{equation}
This matrix can in general be dense, yielding this computation $\mathcal{O}(N \cdot d^3)$ expensive.

There are a variety of methods and architectures one can choose when constructing a NF. To bypass the Jacobian computation expense, we implement the \emph{Block Neural Autoregressive Flow} (BNAF) \citep{decao2019blockneuralautoregressiveflow}, a variation of the neural autoregressive flow (NAF) \citep{huang2018neuralautoregressiveflows}. BNAFs employ a NN architecture where the transformations are applied in blocks, each one generating one dimension of the output at a time, conditioned on the previously generated dimensions. The autoregressive property and  block structure of BNAFs leads to more stable training and mitigated computational expense, which are essential for implementing SVI for high-dimensional problems.

The crux of NAFs that makes them so efficient is that one can construct $q_{\phi}(x)$ such that its Jacobian is lower-triangular, and thus its determinant that must be computed in equation \ref{eqn:flowtransfo} is a simple product:
\begin{equation}
\left| \text{det} \left( \mathbf{J}_{q_{\phi}(z)} \right) \right| = \prod_{i=1}^{d} \frac{\partial f_{i}(z_{i-1})}{\partial z_i} \; .
\end{equation}
The Jacobian is then computed with backpropagation which is $\mathcal{O}(N \cdot d)$ expensive, a clear advantage over the non-autoregressive NF. The typical NAF architecture is a set of functions $f^{(i)}$, where each $f^{(i)}$ can be decomposed into ``conditioners'' $c^{(i)}$ and invertible ``transformers'' $t^{(i)}$
\begin{equation}
    f_{\phi}(x_{<i}) = t_{\phi}^{(i)}(x_i, c_{\phi}^{(i)}(x_{<i})) \; .
\end{equation}
Despite this structure satisfying all the basic needs of a flow, we see that the number of parameters scales quadratically. To bypass this expense, the BNAF structure models each $t_{\phi}^{(i)}$ directly as an NN with no accompanying conditioner. These are all necessary ingredients for a high-dimensional problem like stasis; however, this comes at the expense of some functionality. Despite being theoretically invertible, accessing the inverse of the BNAF, which can be used to compute posterior samples' probability densities, is not currently computationally feasible.

\subsection{Searching for Stasis Theories with SVI}
\label{sec:theoryspaces}

We have now gathered all the essential components to conduct SVI on the stasis simulation, and will use it to sample configurations of rates and abundances that yield stasis, from which we will be able to understand trends of stasis configurations. For this, we employ \texttt{numpyro} \citep{phan2019composable}, a probabilistic programming library that leverages \texttt{jax} for automatic differentiation in the implementation of SVI. Together with \texttt{diffrax}, these packages can utilize GPU computation, significantly speeding up the inference process.

\begin{figure*}[!htbp]
    \centering
    \includegraphics[width=\textwidth]{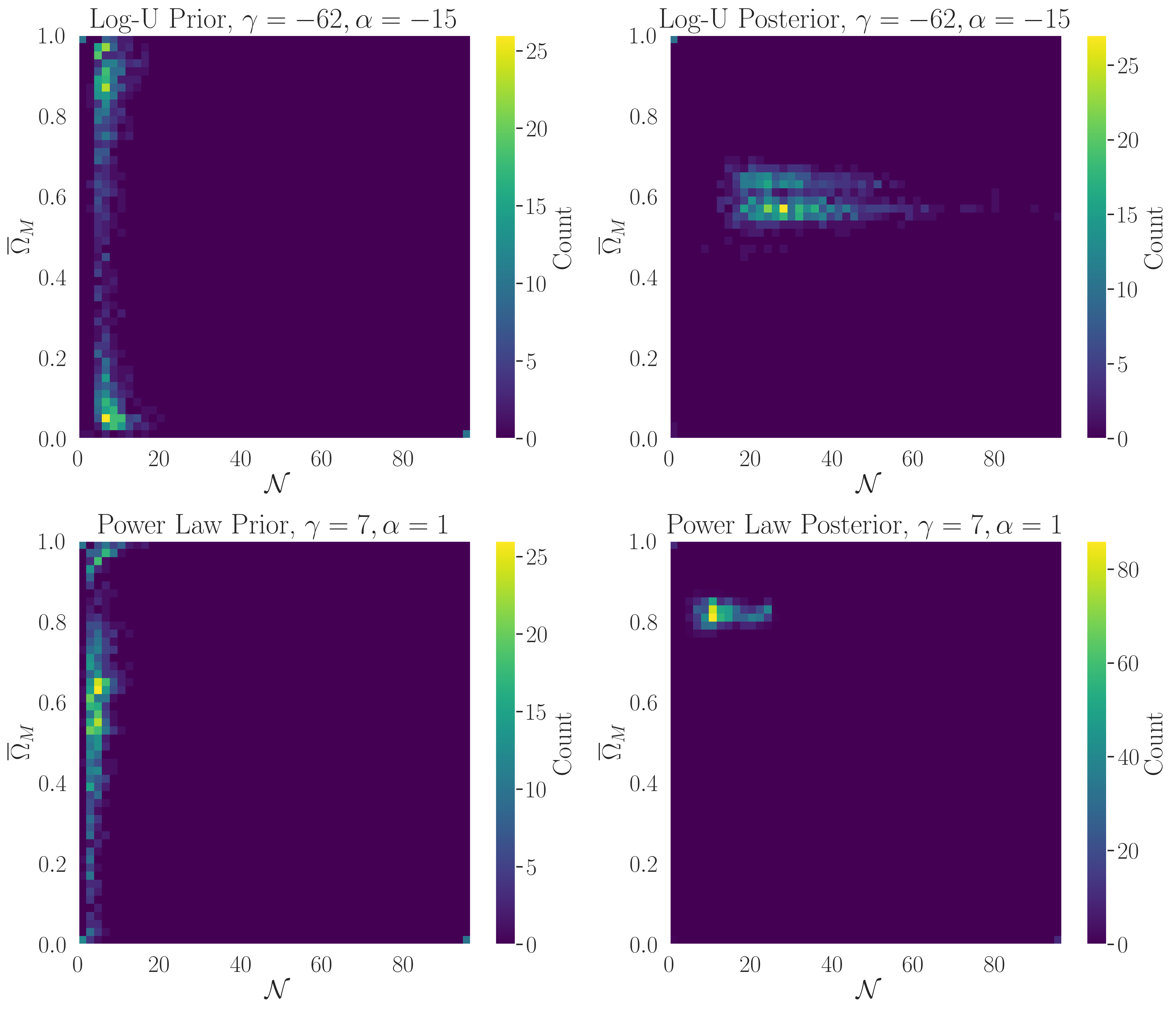}
    \caption{Prior and posterior comparison for choices of power-law prior with $\gammaell \sim \ell^7$ and $\omegaell \sim \ell$ and log-uniform prior with choices of $\gamma = -62$ and $\alpha = -15$ for $N=50$. Each individual heat map is a depiction of 1000 samples and their stasis configuration in $(\mathcal{N},\overline{\Omega}_M)$ space. For both choices of priors, there is a higher degree of stasis in the posterior in both mean and maximum value. The power-law posterior features a mean stasis value of 13.59 $e$-folds and maximum of 25.24 $e$-folds, while the log-uniform posterior has a mean stasis value of 31.06 $e$-folds and maximum of 96.5 $e$-folds.  Over 1000 samples, both priors had $<1 \%$ of samples achieve a stasis epoch of more than 10 $e$-folds, while the power-law and exponential posteriors have $76 \%$ and $99\%$, respectively. While SVI is able to find non-trivial distributions of $\gammaell$ and $\omegaell$ that results in epochs of stasis, it is clear to see the effect of the prior regularization in optimization in the large discrepancy between posterior configurations.}
    \label{fig:stasisheatmap}
\end{figure*}

We now revisit our choice of the surrogate optimization likelihood in equation \ref{eqn:surllhood} instead of the simulator likelihood in equation \ref{eqn:llhood}. The likelihood function is theoretically a normalized PDF that encodes the probability of the observed data given the model parameters. However, in this theoretical particle cosmology setting, there are no direct observations; instead, we aim to understand the input parameter space that produces epochs of stasis. This raises the question: how can SVI be made appropriate here?

\begin{algorithm}
\caption{Searching Stasis with SVI}
\label{alg:algo2}
\begin{algorithmic}[1]
\Require $\phi$ variational parameters, $\theta = \{\gammaell, \omegaell, H^{(0)} \}$ simulation parameters, $p(\Gamma_\ell)$ prior over decay rates, $p(\Omega^{(0)}_\ell)$ prior over initial abundances, $\Gamma_{N-1} / H^{(0)}$ fixed early time decays relative to Hubble constant, $\xi$ early-stopping threshold 
\While{not converged}
    \State $z_0 \sim q_{0}(z)$ \Comment{Sample initial latent variables}
    \State $(\Gamma_\ell, \Omega^{(0)}_\ell) \gets \text{q}_{\phi}(z_0)$ \Comment{Normalizing Flow}
    \State $\Omega_1^{(0)} \leq \Omega_2^{(0)} \leq \ldots \leq \Omega_N^{(0)}$ \Comment{Differentiably sort $\Omega_\ell^{(0)}$}
    \State $\Gamma_1 \leq \Gamma_2 \leq \ldots \leq \Gamma_N$ \Comment{Differentiably sort $\Gamma_\ell$}
    \State $\mathcal{N} \gets \mathcal{S}(\Gamma_\ell, \Omega^{(0)}_\ell, H^{(0)})$ \Comment{Run simulation}
    \State $p(\mathcal{N} | \Gamma_\ell, \Omega^{(0)}_\ell, H^{(0)})$ \Comment{Compute surrogate likelihood}
    \State $\mathcal{L} \gets \text{ELBO}(\phi, \Gamma_\ell, \Omega^{(0)}_\ell, \mathcal{N})$ \Comment{Compute ELBO}
    \State $\nabla_\phi \propto \{ \nabla_{\omegaell}\mathcal{L},\nabla_{\gammaell}\mathcal{L} \}$ \Comment{Backpropagate through stasis simulation}
    \State $\Gamma_\ell \gets \text{clip}(\Gamma_\ell, 0, 1)$ \Comment{Ensure $\Gamma_\ell$ physical}
    \State $\Gamma_{N-1} /  H^{(0)} \gets 0.1 $ \Comment{Enforce initial decays time scale}
    \State $\Delta\phi \propto -\nabla_{\phi}\mathcal{L}( \Gamma_\ell, \Omega^{(0)}_\ell)$ \Comment{Update $\phi$}
    \If{$\mathcal{L}$ does not improve over $\xi$ epochs or \texttt{NaN} loss}
    \State \textbf{break} \Comment{Early-stopping criterion}
    \EndIf
\EndWhile
\end{algorithmic}
\end{algorithm}

The answer lies in the definition of the optimization likelihood given in equation \ref{eqn:surllhood}. 
In a typical Bayesian inference setting, the likelihood function captures the probability of the observed data given the model parameters. This can often be interpreted in terms of residuals between simulation outputs and observed data. \emph{Minimizing} this residual corresponds to \emph{maximizing} the log-likelihood which enters the ELBO, and therefore results in a concrete optimization objective. However, instead of following this traditional approach, we adopt a utility-oriented perspective of the likelihood, treating it purely as an optimization objective. The simulation likelihood given in equation \ref{eqn:llhood} is unsuitable for optimization as it is singular and lacks a well-defined gradient. For this reason, we adopt the optimization likelihood.

A schematic of the SVI pipeline is provided in Figure \ref{fig:svipipeline}, highlighting the four major components of the SVI pipeline: the prior distribution, the variational family, the simulation $\mathcal{S(\theta)}$ and stasis-finder used in the likelihood calculation, and gradient update.  A more detailed description of how we employ SVI in this setting is given in Algorithm 2, where we are again reminded the importance of preserving differentiability. A methodology such as SVI requires that one be differentiable \emph{at every step}, from differentiably sorting $\Omega_\ell^{(0)}$ and $\Gamma_\ell$, to solving the Boltzmann equations while preserving their gradients, and calculating the stasis duration and matter abundance in a differentiable way. This leads to an uninterrupted flow of gradients through the pipeline. 

Conducting these numerical analyses under the Bayesian machine learning paradigm offers both the benefits of expressivity of neural networks with the statistical rigor of Bayesian analysis. We have introduced an alternative formulation of SVI that allows one to operate without direct observations, yet still exploit the properties and benefits of SVI. This approach generally enables the study of otherwise prohibitively difficult high-dimensional parameter spaces, and can be adapted to any physical system that can be simulated. 

\begin{figure*}[!htbp]
    \centering
    \includegraphics[width=\textwidth]{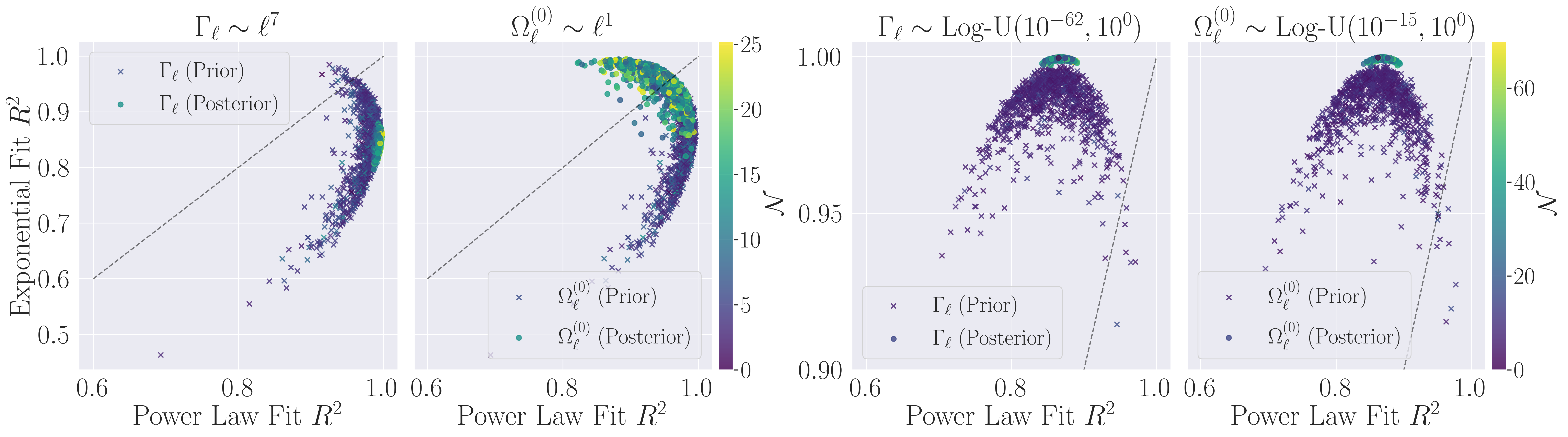}
    \caption{Optimizing stasis with SVI and the ELBO loss (equation \ref{eqn:ELBO}) for log-uniform ($\alpha$ = -15, $\gamma$ = -62) and power-law ($\alpha$ = 1, $\gamma$ = 7) initializations for $N=50$ and $\Gamma_{N-1} / H^{(0)} = 0.1$. A BNAF with two hidden layers and a hidden layer width of 8 neurons was trained for 2000 epochs with Adam optimizer and early-stopping. A batch size of 10 was used in training. (\textbf{Left}) A comparison of model fit for 1000 power-law prior and posterior samples. A flow toward a strictly more power-law posterior is seen for $\gammaell$, with $\omegaell$ becoming more exponential in the posterior, with a comparable mean $R^2$ score for both power-law and exponential fits in the posterior. (\textbf{Right})  A comparison of model fit for 1000 exponential prior and posterior samples. Posteriors for both $\gammaell$ and $\omegaell$ become strictly more exponential. This posterior additionally has a much larger mean stasis $e$-folds and maximum stasis $e$-folds than the power-law prior.}
    \label{fig:r2plotsvi}
\end{figure*}

\subsection{Stasis Results with SVI}

The analyses of the previous sections exhibited a flow in the space of parameters that preferred an exponential model of stasis. Additionally, it was shown that an exponential model had manifestly more robust stasis epochs than the power-law prior corresponding to the model in \citep{Dienes_2022}. Using both sets of priors, it is then natural to consider whether such flows persist when using neural networks to do inference on the full parameter space. 

We run SVI with both priors (power-law and log-uniform) for $N=50$ and $\Gamma_{N-1} / H^{(0)} = 0.1$. We denote the power law prior samples as being drawn from $\omegaell \sim \ell^\alpha$ and $\gammaell \sim \ell^\gamma$. Similarly, we denote the sort-correlated log-uniform samples as being drawn from $\omegaell \sim \text{Log-U}(10^{\alpha}, 10^0)$ and $\gammaell \sim \text{Log-U}(10^{\gamma}, 10^0)$. A numerical prefactor of $\kappa=10$ was used in the surrogate likelihood. We utilize a BNAF with two hidden layers and a hidden layer width of 8 neurons. A batch size of 10 was used for all experiments, corresponding to 10 log-likelihood evaluations per training step. We use Adam \citep{kingma2017adammethodstochasticoptimization} optimizer with an initial learning rate of $\eta = .01$ for 2000 epochs of training with an early-stopping threshold of $\xi = 200$ epochs. All training was done on a single NVIDIA A100-80GB GPU.

We find that the power-law prior was susceptible to a mode collapse of either completely matter dominated or radiation dominated stasis configurations using the likelihood defined in equation \ref{eqn:surllhood}. To address this, an additional constraint on the likelihood was imposed, similar to the regularization condition when just using gradients in the differentiable simulation, shown in equation \ref{eq:gradcondition}. The experiments with the log-uniform prior did not exhibit this mode collapse.

We find that SVI is able to learn distributions of $\gammaell$ and $\omegaell$ that result in epochs of stasis for both choices of priors, as shown in the $\mathcal{N}-\overline{\Omega}_M$ space heat maps for 1000 posterior samples in Figure \ref{fig:stasisheatmap}. Samples from the power-law prior exhibit a mean stasis value of 4.61 $e$-folds and maximum of 16.51 $e$-folds, which after optimization is 13.59 $e$-folds and 25.24 $e$-folds in the posterior. The posterior/prior discrepancy of the exponential prior experiment is much more drastic, with prior samples exhibiting a mean of 7.46 $e$-folds and maximum of 19.38 $e$-folds, which is 31.06 $e$-folds and 96.49 $e$-folds in the posterior. While the SVI optimization has worked in both scenarios, it is clear by the discrepancies in posterior configurations the effect that the choice of prior has in the optimization, as shown in equation \ref{eqn:ELBO}. This was indeed expected from the numerical study of prior configurations shown in Figure \ref{fig:heatmapcomparison}; it is therefore interesting to study the model dependence of posterior samples and their flow from model dependence in their priors.

The flow in the space of parameters of $\gammaell$ and $\omegaell$ for both experiments is shown in Figure \ref{fig:r2plotsvi}. For the power-law prior, it is seen that $\gammaell$ and $\omegaell$ are more confidently fit with a power-law model, as expected. After training, there is a clear drift towards a more power-law model in $\gammaell$. Interestingly, it is only $\omegaell$ that displays a drift towards an exponential model with the $\omegaell$ posterior featuring a mean fit value of $R^2 = 0.95$ for both power-law and exponential, with the mean power-law $R^2$ value decreasing from $R^2 = 0.99$ in the prior and the mean exponential fit $R^2$ value increasing from $R^2 = 0.87$ in the prior. This flow of parameters is contrary to the result in which just the differentiable simulation was used (Figure \ref{fig:r2plot}), where both $\gammaell$ and $\omegaell$ drift towards an exponential.  The posterior in this case is a qualitative hybrid of both models of stasis, and even in such conditions the emergence of stasis epochs is possible. Stasis epochs are not physically exclusive to single models, whether they be power-law or exponential in parameters. 

For the log-uniform prior, both $\gammaell$ and $\omegaell$ posteriors display clear preference towards an exponential model with a mean score of $R^2 = 1$ for an exponential fit. The mean power-law fit $R^2$ score is invariant after optimization, but it is seen in Figure \ref{fig:r2plotsvi} that the scatter has significantly decreased.

These results highlight that in a properly Bayesian setting, both a power-law and exponential model for $\gammaell$ and $\omegaell$ are valid solutions when conditioning on producing epochs of stasis. The drift towards an exponential model of $\omegaell$ even when having a power-law prior can indicate that completely power-law distributions of $\gammaell$ and $\omegaell$ are more sparse in the full parameter space. Additionally, the discrepancy in the \emph{statistics} of stasis in the two posteriors illuminate a sharp difference in the persistence of stasis epochs for the two types of models -- a mean difference of $18$ $e$-folds.

It could be criticized that the flow of the posterior towards an exponential model is purely due to the choice of prior. We see that this is not strictly the case as with the $\omegaell$ parameter flow for the power-law prior experiment, in which the parameter flow goes distinctly \emph{against} the prior regularization in the ELBO. Thus, the prior plays an important role in the ELBO, but SVI can lead to results that are qualitatively at odds with the prior, due to the condition intrinsic to the posterior. 

\section{Models of Stasis}
\label{sec:models}

Thus far, we have remained as model-agnostic about stasis as possible, choosing only the prior distribution on rates and abundances for both our gradient ascent and Bayesian inference analyses. In both analyses, optimization was chosen to optimize stasis, with a clear preference for exponential models over power-law models. 

In this section we wish to study an exact exponential model both analytically and numerically, compare it to exact power-law models, and provide a preliminary analysis of stasis in the String axiverse; more thorough analysis will be reserved for future work. By ``exact" in both contexts, we mean a model with fixed formulae for $\gammaell$ and $\omegaell$ and no noise, as opposed to the intrinsically noisy draws of previous sections.

\subsection{An Exact Exponential Model of Stasis}
Motivated by our numerical results, we introduce an exponential model of stasis that yields parametrically more stasis than power-law models, as a function of the number of species. We will  show that such a model obeys the dynamical equations of stasis and will result in robust, finite epochs of stasis. We additionally derive algebraic relations with exact predictions for $\mathcal{N}$ and $\overline{\Omega}_M$ from our model.i

We begin my parameterizing this model as
\begin{equation}
    \gammaell =  \Gamma_N e^{\gamma(\ell - N)}, \qquad \omegaell = \Omega_N^{(0)} e^{\alpha(\ell - N)}
\end{equation}
for a spectrum of $N$ species indexed by $\ell$ and general scaling factors $\alpha$, $\gamma$, and $\Gamma_N$. As in the original model of stasis, $\alpha$ and $\gamma$ control the hierarchies in abundances and rates, this time with exponential dependence. We are also reminded that $\omegaell(t^{(0)}) = 1$, which requires the overall normalization factor $\Omega_N^{(0)} = \left[\sum_{\ell = 1}^{N} e^{\alpha(\ell - N)}\right]^{-1} $. However, in contrast with the model in \citep{Dienes_2022}, where the fundamental object was the mass spectrum $m_\ell$, we choose to parameterize this model of stasis directly with the decay rates and abundances. We are thus ready to define the five parameter model inspired by an exponential:
\begin{equation}
    \{\alpha, \gamma, \Gamma_N, \Omega_N^{(0)}, t^{(0)} \} \; ,
\end{equation}
with which shall proceed to show obeys the dynamical equations required of a stasis epoch.

We begin by reminding ourselves that a model of stasis must satisfy the constraint equations given in equations \ref{eqn:stasiscondition1} and \ref{eqn:stasiscondition2}. Focusing on the time evolution of the total $\omegaell$, we get 
\begin{equation}
\label{eqn:omegaellsum}
    \sum_\ell \Omega_\ell(t) = \Omega_\ell^{(0)}h(t^{(0)}, t) \sum_{\ell = 1}^{N} e^{\alpha(\ell - N)}e^{-\Gamma_N(t - t^{(0)}) e^{\gamma(\ell-N)}} \; ,
\end{equation}
where we have used the result from equation \ref{eqn:fiducialomega}, but replaced the time dependence with a more general net gravitational redshift factor $h(t^{(0)}, t)$ such that there is no assumption that we are in a period of stasis. $h(t^{(0)}, t)$ invariant across all species (i.e. is independent of $\ell$); a more in-depth discussion of this $h$-factor can be found in \citep{Dienes_2022}. Moving to the $N \rightarrow \infty$ limit, we simplify this calculation by transforming the sum into an integral
\begin{align}
\label{eqn:continuum}
    \sum_{\ell = 1}^{N} &e^{\alpha(\ell - N)}e^{-\Gamma_N(t - t^{(0)}) e^{\gamma(\ell-N)}} \nonumber \\ &\rightarrow \int_0^{\infty} e^{\alpha(\ell - N + 1)}e^{-\Gamma_N(t - t^{(0)}) e^{\gamma(\ell-N + 1)}} d \ell \; ,
\end{align}
which can be computed by invoking the Gamma function identity
$\Gamma(k)/b^k = \int_0^\infty z^{k-1}e^{-bz} dz,$ yielding
\begin{align}
    \sum_\ell \Omega_\ell(t) = \frac{\Omega_N^{(0)}}{\gamma} &\Gamma \left(\frac{\alpha}{\gamma}\right) h(t^{(0)}, t) \nonumber \\
    &\times \left[\Gamma_N (t-t^{(0)})\right]^{-\alpha / \gamma} \; .
\end{align}
We can solve for the quantity $\sum \gammaell \Omega_\ell(t)$ similarly, resulting in
\begin{align}
\label{eqn:modelstasisequation2}
    \sum_\ell \Gamma_\ell \Omega_\ell(t) = \frac{\Omega_N^{(0)} \Gamma_N}{\gamma} &\Gamma \left(\frac{\alpha}{\gamma} + 1\right) h(t^{(0)}, t) \nonumber \\
    &\times \left[\Gamma_N (t-t^{(0)})\right]^{-(\alpha / \gamma + 1)} \; .
\end{align}
It is lastly straightforward to compute the ratio of these conditions, where we invoke the Gamma function identity $\Gamma(z+1) / \Gamma(z) = z$,
\begin{eqnarray}
\label{eqn:modelstasiscondition1}
    \frac{\sum_\ell \gammaell \Omega_\ell(t)}{\sum_\ell \Omega_\ell(t)} = \frac{\alpha}{\gamma} \left( \frac{1}{(t - t^{(0)})} \right) \; .
\end{eqnarray}
We see that this result is power-law in the difference $(t-t^{(0)})$, dissimilar to the $t^{-1}$ dependence from the stasis dynamical equations in equations \ref{eqn:stasiscondition1} and \ref{eqn:stasiscondition2}. Indeed, this discrepancy also arises in the power-law model of stasis. Qualitatively, this means that the stasis epoch must emerge some time after the initial species production time, when $t \gg t^{(0)}$ and hence $(t - t^{(0)})^{-1} \approx t^{-1}$. This is when the edge effects, most apparent when $\Gamma_{N} / H^{(0)} \gg 1$, have died away. This further indicates that our model produces \emph{finite} epochs of stasis, with a natural beginning of the stasis epochs emerging due to the presence of edge effects, and an end when all species have decayed.

With this result in hand, we can proceed to compare coefficients in \ref{eqn:stasiscondition2} and \ref{eqn:modelstasiscondition1}, in which we find
\begin{equation}
\label{eqn:parametercondition}
    \frac{\alpha}{\gamma} = \frac{2 (1 - \overline{\Omega}_M)}{4 - \overline{\Omega}_M} \; .
\end{equation}
We can further invert this to obtain the prediction for $\overline{\Omega}_M$ during stasis in terms of our model parameters, resulting in
\begin{equation}
    \overline{\Omega}_M = \frac{2 (\gamma - 2 \alpha)}{2\gamma - \alpha} \; ,
\end{equation}
from which we can also find a parameter restriction for $\alpha$ and $\gamma$ by enforcing $0 < \overline{\Omega}_M < 1$:
\begin{equation}
    0 < \alpha < \frac{\gamma}{2} \; .
\end{equation}
With this result, we can identify the model parameters that result in matter-radiation equality (MRE). Solving equation \ref{eqn:parametercondition} for $\overline{\Omega}_M = 1/2$, we get the condition
\begin{equation}
    \frac{\alpha}{\gamma} = \frac{2}{7}
\end{equation}
for MRE. We will study this result numerically and further compare it to the power-law model of MRE.

It remains now to check that the constraints for $\sum_\ell \Omega_\ell(t)$ (equation \ref{eqn:stasiscondition1}) and $\sum_\ell \Gamma_\ell \Omega_\ell(t)$ (equation \ref{eqn:stasiscondition2}) are individually satisfied. To that end, without loss of generality we can simply show that the condition in equation \ref{eqn:stasiscondition1} is satisfied, as we have already checked the condition of their ratio. We begin by operating under $t \gg t^{(0)}$, for which we have
\begin{align}
    h(t^{(0)}, t) &= h(t^{(0)}, t_*) h(t_*, t) \nonumber \\
    &= h(t^{(0)}, t_*) \left( \frac{t}{t_*} \right)^{2 - 6/(4 - \overline{\Omega}_M)} \; ,
\end{align}
where we have inserted the factor for $h(t_*,t)$ from equation \ref{eqn:fiducialomega}, and $t_* \gg t^{(0)}$ is some fiducial time much later than the initial species production. Inserting this into equation \ref{eqn:modelstasiscondition1}, we arrive at
\begin{align}
    \sum_\ell \Omega_\ell(t) = \frac{\Omega_N^{(0)}}{\gamma} &\Gamma \left(\frac{\alpha}{\gamma} \right) h(t^{(0)}, t_*) \left(\frac{t}{t_*}\right)^{2 - 6/(4 - \overline{\Omega}_M)} \nonumber \\
    &\times \left[\Gamma_N(t - t^{(0)})\right]^{-\alpha/\gamma} \; ,
\end{align}
from which we see using the result in equation \ref{eqn:parametercondition} and considering that $t \gg t^{(0)}$, the time dependence falls out, ensuring that we are in a period of stasis in which $\Omega_M \equiv \sum_\ell \Omega_\ell(t)$. Our exponential model of stasis is thus consistent with the conditions for stasis as defined from their corresponding dynamical equations.

\begin{figure}
    \centering
    \includegraphics[width=.95\columnwidth]{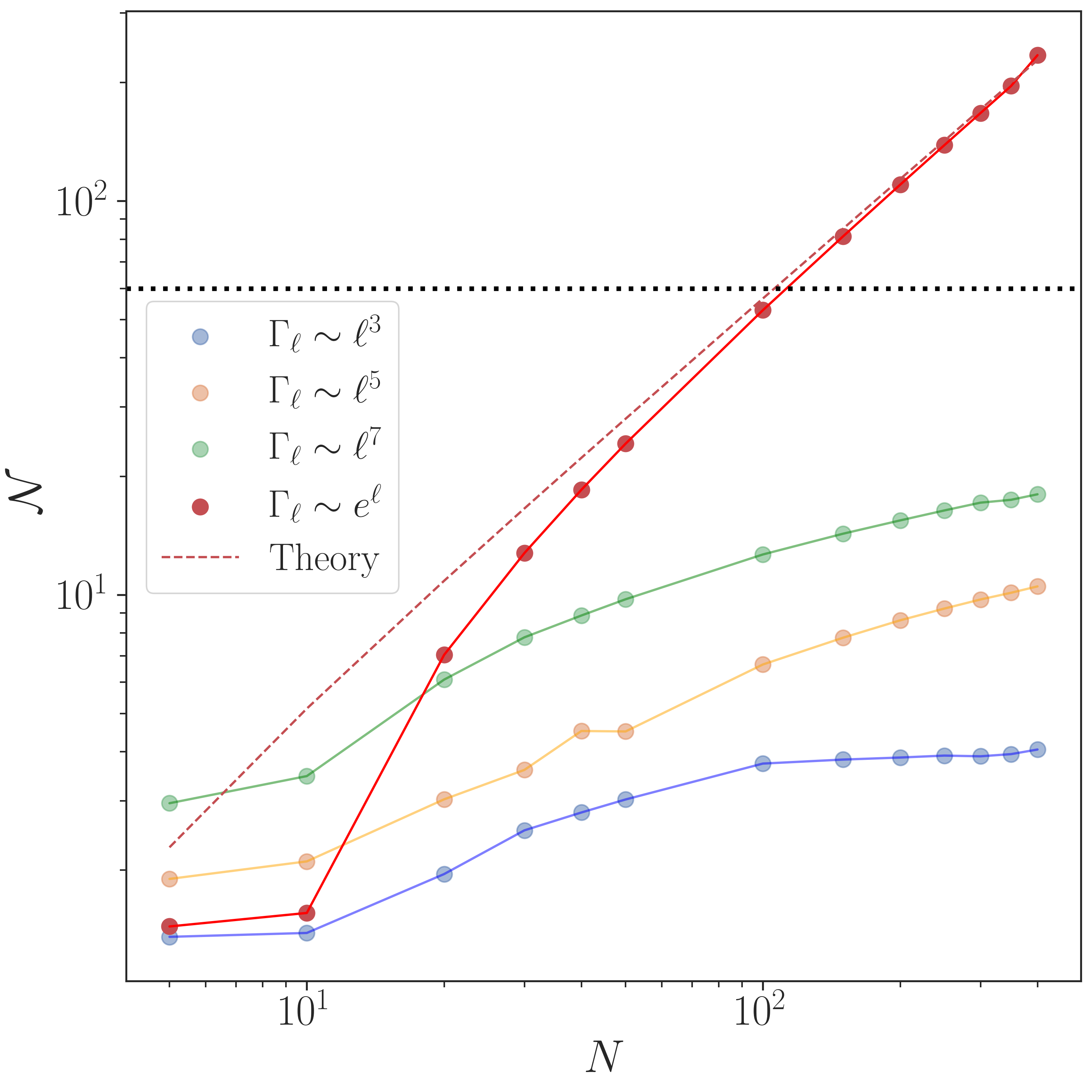}
    \caption{Comparison between exponential model of stasis with $\gamma = 1$, $\alpha = 2/7$, and $\Gamma_N = 0.01$ yielding matter-radiation equality, and the power-law model of stasis with $\alpha=\delta=1$. The theoretical prediction of $e$-fold scaling for the exponential model from equation \ref{eqn:efoldtheory} is shown in the red dashed line, in which we see that in the $N \rightarrow \infty$ limit there is exact agreement between numerical data and the theoretical prediction. MRE for the power-law model with $\gamma=7$ is shown in green, with the power-law model exhibiting logarithmic scaling with $N$ and the exponential model exhibiting linear scaling with $N$ as $N \rightarrow \infty$. It is also seen that more than 60 $e$-folds of MRE (black dashed line) is achieved with just $N \sim 100$ species for the exponential model.}
    \label{fig:efoldcomparison}
\end{figure}

We are now in a position to estimate exactly the number of $e$-folds of stasis expected from a configuration of our model, given by:
\begin{align}
\label{eqn:efoldtheory}
    \mathcal{N} \equiv \log \left[ \frac{a(t = \tau_1)}{a(t = \tau_{N})}\right] &= \frac{2}{4 - \overline{\Omega}_M} \log \left( \frac{\Gamma_{N} }{\Gamma_1} \right) \nonumber \\
    &= \frac{2 \gamma}{4 - \overline{\Omega}_M} \left( N -1 \right) \; ,
\end{align} 
where we have used the result for the time evolution of $a(t)$ in equation \ref{eqn:scalefactor}. 

\begin{figure*}[!htbp]
    \centering
    \includegraphics[width=\textwidth]{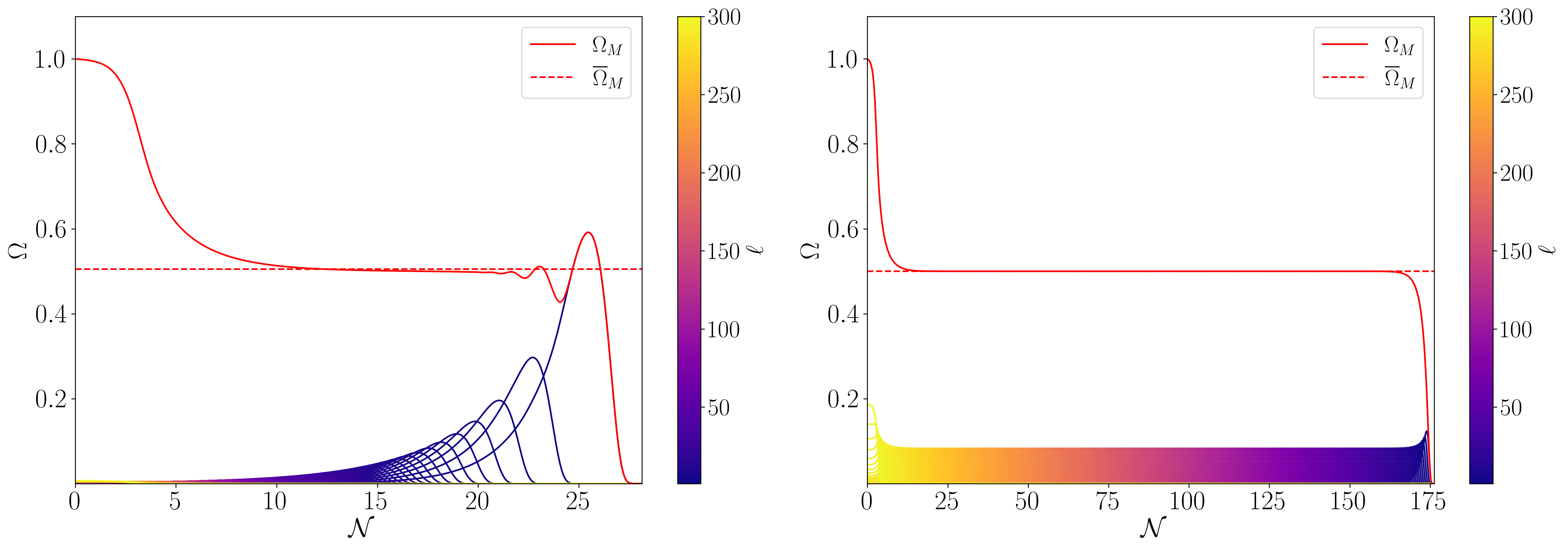}
    \caption{Comparison of matter-radiation equality (MRE) between the power-law (left) and exponential (right) models of stasis, both for $N=300$ species. MRE in the power law model from \citep{Dienes_2022} corresponds to $\alpha = \delta = 1$ and $\gamma = 7$ and with $\Gamma_{N-1} / H^{(0)} = 0.01$. MRE in the exponential model corresponds to $\gamma=1$, $\alpha = 2/7$, and $\Gamma_N/H^{(0)} = 0.01$. We see that the power-law model achieves a stasis configuration of $\sim 17$ $e$-folds with 300 species, while the exponential model achieves $\sim 165$ $e$-folds, demonstrating the qualitative difference in $e$-folds between the two models, attributable to different scaling of $e$-folds with $N$.}
    \label{fig:manyefolds}
\end{figure*}

We see that the exponential model of stasis yields parametrically more e-folds than the power-law model, which exhibited $\mathcal{N} \sim \log(N)$ scaling. A numerical comparison of $e$-fold data and the theoretical prediction for $\gamma=1$, $\alpha = 2/7$, and $\Gamma_N = 0.01$ yielding a stasis epoch with matter-radiation equality (MRE), is shown in Figure \ref{fig:efoldcomparison}. Further, see Figure \ref{fig:manyefolds} for a comparison of stasis plots with $N=300$ species in the exponential and power-law model. As expected, there is excellent agreement between data and theory in the limit $N \rightarrow \infty$. The figure also illustrates that more than 60 $e$-folds of MRE can be achieved with just 100 species in the exponential model. This result shows that a relatively small number of species is needed to achieve inflation-level $e$-folding, which may have important phenomenological implications.

We additionally compare $e$-fold scaling with the power-law model of stasis in Figure \ref{fig:efoldcomparison}, where $\alpha = \delta = 1$ and $\gamma$ is being varied. The figure demonstrates the qualitative difference in $e$-folds for stasis epochs between the two models, even with strong power-law scaling of $\gammaell \sim \ell^7$ corresponding to MRE. Specifically, the exponential model achieves the same $e$-folds of stasis with just $N \sim 50$ species, compared to the 400 species required for $\gammaell \sim \ell^7$. The logarithmic scaling in the power-law model and the linear scaling at high $N$ for the exponential model are both evident. Therefore, while both models produce epochs of stasis, it is clear that the exponential model results in  longer epochs of stasis. 

We lastly recall that in \citep{Dienes_2022}, it was shown that a stasis state is a \emph{global attractor} of the dynamical system. This was demonstrated to be true for the power-law model, and we can proceed to show minimally that it holds for the exponential model as well. In doing so, it is sufficient to see that for the exponential model
\begin{align}
\sum_\ell \Gamma_\ell \Omega_\ell &= \left( \frac{\alpha}{\gamma} \right) \frac{\Omega_M}{t - t^{(0)}} \nonumber \\
&= \left[ \frac{2(1 - \overline{\Omega}_M)}{4 - \overline{\Omega}_M}\right] \frac{\Omega_M}{t - t^{(0)}} \; ,
\end{align}
where we see that this is exactly the result of equation 4.1 in \citep{Dienes_2022} up to a constant factor.  It then follows that the subsequent analyses showing that the stasis state is an attractor also apply to our exponential model of stasis.

\subsection{Stasis and the String Axiverse}
\label{sec:axiverse}

Axions arise readily in string theory via the dimensional reduction of higher-form
gauge fields; see, e.g., \cite{Svrcek:2006yi,Arvanitaki:2009fg}. Notably, they need not be the QCD axion or axions that couple
to any dark gauge sector. Accordingly, string theory axions can be relevant in a number of phenomenological or cosmological roles,
including as the inflaton \cite{Dimopoulos:2005ac,Kim:2004rp,Silverstein:2008sg,McAllister:2008hb}, as the reheaton \cite{Halverson:2019kna}, as the QCD axion \cite{Svrcek:2006yi, Conlon:2006tq,Demirtas:2021gsq,Gendler:2024adn}, as particles that couple to photons \cite{Halverson:2019cmy,Gendler:2023kjt}, and more. The largest number of axions in a known Type IIB / F-theory compactification is 181,820 \cite{wang2020elliptic} in a single geometry, whereas \cite{Halverson:2017ffz,Taylor:2017yqr} and \cite{Kreuzer:2000xy,Demirtas:2018akl} give rise to large ensembles of geometries that typically have thousands or hundreds of axions, respectively.

Could axions give rise to stasis? Features that enable robust periods of stasis include a spectrum of $N$ particles that may decay to radiation and non-trivial hierarchies in the decay rates. The former is clearly satisfied in many corners of the string landscape, but in lieu of a detailed analysis we would like to comment on an essential feature that affects rates: in many string constructions, axion masses are generated non-perturbatively and the $\ell^\text{th}$ axion masses appears schematically as 
\begin{equation}
m_\ell \propto e^{-c\, T_\ell}
\end{equation}
where $T_\ell$ is the volume of the cycle wrapped by the string instanton that generates the mass. Decays to radiation often occur via dimension five operators, in which case we have 
\begin{equation}
\Gamma_\ell \propto e^{-3c\, T_\ell} \qquad \Omega_\ell^{(0)} \propto e^{-\alpha c \, T_\ell},
\end{equation}
where $\alpha$ depends on the production mechanism as discussed in \cite{Dienes_2022}. Though the rates and abundances have power-law dependence on masses, the exponential dependence of the non-perturbative mass causes it to depend exponentially on the cycle volumes, which are themselves sensitive to the details of moduli stabilization. However, due to the non-perturbative effects there is a clear opportunity for exponential hierarchies, and in fact in string theoretic constructions of the models similar to the Standard Model, the exponential dependence on $T_\text{QCD}$ sets $\Lambda_\text{QCD}$ exponentially below the Planck scale. Furthermore, if moduli stabilization distributes $T_\ell$ uniformly, then the rates and abundances are log-uniform and can achieve many $e$-folds of stasis, as we have shown.

\subsection{Stasis and the Emergent String Conjecture}

The Emergent String Conjecture \cite{lee2022emergent} is a generalization of the Swampland Distance Conjecture \cite{ooguri2007geometry} in which a tower of either Kaluza-Klein or string states
\begin{equation}
m_\ell(\phi) = m_\ell(\phi_0) e^{-\lambda d(\phi,\phi_0)}
\end{equation}
becomes exponentially light upon going a distance $d(\phi,\phi_0)$ from $\phi_0$ toward $\phi$ near the boundary of moduli space. The towers still have power-law masses, and therefore the number of $e$-folds of stasis satisfies 
\begin{equation}
\mathcal{N} \propto \log(N),
\end{equation}
but the exponential suppression of the lowest mass state yields an exponentially large number of states below a fixed cutoff. We therefore have 
\begin{equation}
\mathcal{N} \propto \frac{\lambda d(\phi,\phi_0)}{\delta},
\end{equation}
where $m_\ell \propto \ell^\delta$. We therefore expect stasis to become increasingly important as the boundary of moduli space is approached, potentially leading to experimental constraints or observational consequences.

\section{Discussion}
\label{sec:conclusions}

In this paper, we have studied $M\rightarrow \gamma$ cosmological stasis. In this scenario, a tower of matter states initially dominates the energy density of the universe, but subsequent decays to radiation balance against Hubble expansion, leading to a cosmological epoch with approximately constant matter and radiation abundances.  

A central result of this paper is a new exponential model of stasis that was motivated by our numerical approach, searching the full $2N$-dimensional parameter space of decay rates $\gammaell$ and initial abundances $\omegaell$ using gradients from a differentiable Boltzmann solver, as well as stochastic variational inference with neural networks. This model leads to a longer duration of stasis in $e$-folds relative to a power-law model and could be motivated by non-perturbative effects in string theory. We elaborate on and review our main results in what follows.

A stasis epoch could have significant cosmological implications \citep{Dienes_2022}, potentially affecting dark matter production, large-scale structure formation, and the overall understanding of cosmological evolution and the age of the Universe. Of particular interest for $M \rightarrow \gamma$ stasis is the possibility of stasis epochs occurring after a matter-dominated epoch at the end of inflation or after a later matter-dominated epoch prior to nucleosynthesis. In the first case, the decay of the inflaton can instead source $\phi_\ell$ states. The hierarchical decays of those states during stasis would then be the source of inflationary reheating, with the universe entering radiation domination with the conclusion of the decays. In the second case, the tower of matter fields can lead to an early matter dominated epoch prior to BBN, but well after inflation, the decays of which give rise to stasis. There are a number of potential phenomenological implications of such an epoch, but a stasis state is an attractor regardless of the arena in which $\phi_\ell$ are produced.

\bigskip
 In this work, we have taken a model-agnostic  data-driven approach to studying theories of stasis, employing analytic methods only in the final stages. Such an analysis is complementary to a model-driven approach, as e.g. in the
initial model of stasis that was inspired by Kaluza-Klein excitations, motivating a mass spectrum that follows a power-law in the species index. Together, the model-driven and model-agnostic approaches demonstrate that stasis is a very general phenomenon.

Our methodology began with constructing a differentiable Boltzmann solver, which is the crux of our approach. We maximized the number of $e$-folds of stasis by following gradients in our differentiable simulation with model-agnostic (aside from the prior) random initializations of $\gammaell$ and $\omegaell$. Such stasis-maximizing trajectories through the space of rates and abundances  motivated the study of log-uniform distributions, which we compared to power-law distributions by taking random samples and solving the Boltzmann equations. This comparison provided an initial indication of the significant discrepancy in the duration of stasis epochs between the two models. With distributional priors established, we employed SVI with normalizing flows for a Bayesian analysis of the full, high-dimensional parameter space; there are some essential differences from standard SVI that we discuss in the text. From the posteriors, we again saw that stasis maximization generally prefers log-uniform distributed $\omegaell$ and $\gammaell$, corresponding to a model that is exponential in the species index $\ell$. This motivates a hybrid power-law / exponential model beyond the strict power-law or exponential models we have studied, further indicating the generality of stasis.

The culmination of our numerical analyses is the analytic exponential model of stasis studied in Section \ref{sec:models}. Although our study of stasis does not assume specific physical mechanisms to source the $\phi_\ell$, we find in Section \ref{sec:axiverse} that such a model of stasis could be naturally realized within the string axiverse or near the boundaries of moduli space in accord with the emergent string conjecture.  This new model of stasis is also an attractor.

In addition to the analytic model it inspired, our numerical methodology allows us to conclude with the following observations:
\begin{itemize}
\item Deviations from strict stasis (e.g., $\epsilon$-stasis that could have small oscillations), which can still be phenomenologically relevant, indicate that $\omegaell$ and $\gammaell$ drawn from certain families of distributions can result in epochs of stasis.
\item Stasis epochs can arise without strictly correlated rates and abundances, with the species participating in stasis having abundances increasing with decay rates.
\item There is a numerical preference for an exponential model when maximizing stasis; this was observed in both gradient ascent and SVI experiments.
\item An exponential model of rates and abundances yields more $e$-folds of stasis than the power-law model, with both analytics and numerics demonstrating that $\mathcal{N} \propto N$ in the exponential model whereas $\mathcal{N}\propto \log{(N)}$ in the power-law model. 
\end{itemize}
In our opinion, these model-agnostic numerical analyses together with model-driven analytic considerations together make stasis a very compelling phenomenon from a theoretical perspective. The generality of the alternative cosmological histories implies that in complex cosmologies, such as those in string theory, one should not make assumptions about the history and instead one must simply solve the Boltzmann equations.

In future work, it would be interesting to study other flavors of stasis, notably $\Lambda \rightarrow \gamma$, $\Lambda \rightarrow M$, and triple stasis with $\Lambda \rightarrow M \rightarrow \gamma$, using the numerical techniques presented in this paper. To that end, extending the differentiable methodology to include other energy pumps, such as the overdamped/underdamped transition described in \citep{dienes2023stasis} used to model $\Lambda \rightarrow M$, would be essential. From a numerical perspective, these transitions can be challenging to implement due to their instantaneous (i.e., non-differentiable) nature; however, one can consider a family of numerical approximations (e.g., \texttt{tanh}, \texttt{sigmoid}) to model such transitions in a differentiable way. 

It would also be interesting to change the optimization objective in future work. All of our numerical analyses were aimed at ensuring robust periods of stasis by optimizing the number of stasis $e$-folds $\mathcal{N}$. Of course, this optimization is different from being physically optimal, which depends crucially on theory priors, as well as phenomenological and cosmological viability. The latter considerations motivate different optimization objectives that could easily be implemented by adapting our open-source code \githubmaster, provided that the objective is implemented in a differentiable way. Doing so would open up new avenues for different physical studies of stasis.

We have seen that studies of theories with  high-dimensional parameter spaces using differentiable simulators, despite being arduous work, can be essential to understanding the physics. Specifically, we have demonstrated the power of gradients from these simulators to direct flows in parameter space, both with and without neural networks. The methodology presented here is flexible and easily adaptable. Indeed, these techniques, in conjunction with neural networks, can provide profound physical insights, motivate new physical models, and have far-reaching implications in various scientific fields, including those beyond particle cosmology.

\acknowledgements 
The computations in this paper were run on the FASRC Cannon cluster supported by the FAS Division of Science Research Computing Group at Harvard University. We thank Siddharth Mishra-Sharma and Carolina Cuesta-Lazaro for initial guidance on SVI, Keith Dienes for discussions regarding stasis, and the authors of \cite{Dienes_2022} for comments on a draft of this manuscript. This work is supported by the National Science Foundation under CAREER grant PHY-1848089 and
Cooperative Agreement PHY-2019786 (The NSF AI Institute for Artificial Intelligence and Fundamental
Interactions).

\newpage
\bibliography{stasis}

\end{document}